\documentclass[fleqn]{2020SCGE}

\usepackage{graphicx}
\usepackage{multirow}
\usepackage{hyperref}

\newcommand{\kmsmpc}{\>{\rm km}\,{\rm s}^{-1}\,{\rm Mpc}^{-1}}

\newcommand{\rmd}{{\rm d}}
\newcommand{\rp}{r_{\rm p}}

\usepackage{color}

\begin{document}

\ensubject{subject}

\ArticleType{Article}
\SpecialTopic{SPECIAL TOPIC: }
\Year{2023}
\Month{xxx}
\Vol{xx}
\No{x}
\DOI{xx}
\ArtNo{000000}
\ReceiveDate{xxx xxx, 2023}
\AcceptDate{xxx xxx, 2023}

\title{DESI Legacy Imaging Surveys Data Release 9: Cosmological Constraints from Galaxy Clustering and Weak Lensing using the Minimal Bias Model}{}

\author[1,2,6]{Haojie Xu}{haojie.xu@shao.ac.cn}
\author[1,2]{Hekun Li}{hekun\_lee@sjtu.edu.cn}
\author[1,2]{Jun Zhang}{}
\author[1,2,3]{Xiaohu Yang}{xyang@sjtu.edu.cn}
\author[1,2,3]{Pengjie Zhang}{}
\author[1,4]{Min He}{}
\author[1,2]{Yizhou Gu}{}
\author[1,2]{\\Jian Qin}{}
\author[1,2,5]{Zhaozhou Li}{}
\author[1,2]{Antonios Katsianis}{}
\author[1,6]{Ji Yao}{}
\author[1,2]{Zhaoyu Wang}{}
\author[1,2]{Yirong Wang}{}
\author[7]{Liping Fu}{}

\AuthorMark{H. Xu}

\AuthorCitation{Xu, H. et al.}

\address[1]{Department of Astronomy, Shanghai Jiao Tong University, Shanghai {\rm 200240}, China}
\address[2]{Key Laboratory for Particle Astrophysics and Cosmology (MOE)/Shanghai Key Laboratory for Particle Physics and Cosmology , 
Shanghai {\rm 200240}, China}
\address[3]{Tsung-Dao Lee Institute, Shanghai Jiao Tong University, Shanghai {\rm 201210}, China}
\address[4]{CAS Key Laboratory of Optical Astronomy, National Astronomical Observatories, Chinese Academy of Sciences, Beijing {\rm 100101}, China}
\address[5] {Centre for Astrophysics and Planetary Science, Racah Institute of Physics, The Hebrew University, Jerusalem, {\rm 91904}, Israel}
\address[6]{Shanghai Astronomical Observatory, Chinese Academy of Sciences, Shanghai 200030, China}
\address[7]{Shanghai Key Lab for Astrophysics, Shanghai Normal University, Shanghai {\rm 200234}, China}

\abstract{
We present a tentative constraint on cosmological parameters $\Omega_m$ and $\sigma_8$ from a joint analysis of galaxy 
clustering and galaxy-galaxy lensing from DESI Legacy Imaging Surveys Data Release 9 (DR9), 
covering approximately 10000 square degrees and spanning the redshift range of 0.1 to 0.9. 
To study the dependence of cosmological parameters on lens redshift, we divide lens galaxies into seven approximately volume-limited samples, each with an equal width in photometric redshift.
To retrieve the intrinsic projected correlation function $w_{\rm p}(r_{\rm p})$ from the lens samples, we employ a novel method to account for redshift uncertainties. 
Additionally, we measured the galaxy-galaxy lensing signal $\Delta\Sigma(\rp)$ for each lens sample, using source galaxies selected from the shear catalog by applying our {\tt Fourier\_Quad} pipeline to DR9 images. 
We model these observables within the flat $\Lambda$CDM framework, employing the minimal bias model. 
To ensure the reliability of the minimal bias model, we apply conservative scale cuts: $r_{\rm p} > 8$ and $12 ~h^{-1}{\rm Mpc}$, for $w_{\rm p}(r_{\rm p})$ and $\Delta\Sigma(\rp)$, respectively.
Our findings suggest a mild tendency that $S_8 \equiv \sigma_8 \sqrt{\Omega_m/0.3} $ increases with lens redshift, although this trend is only marginally significant. 
When we combine low redshift samples, the value of
$S_8$ is determined to be $0.84 \pm 0.02$, consistent with the Planck results but significantly higher than the 3$\times$ 2pt analysis by 2-5$\sigma$.
Despite the fact that further refinements in measurements and modeling could improve the accuracy of our results, the consistency with standard values demonstrates the potential of our method for more precise and accurate cosmology in the future.}



\maketitle


\begin{multicols}{2}

\section{Introduction}\label{sec:intro}

The spatial distribution of galaxies contains crucial information about the nature of our Universe. Over the past decades, cosmological models have been constrained from large redshift and imaging surveys using a combination of galaxy clustering and galaxy-galaxy lensing \cite{Cacciato2013, Mandelbaum2013, Leauthaud2017, Abbott2018, Lange2019b, Singh2020}, cosmic shear \cite{Fu2014, Jee2016, Hildebrandt2017, Hikage2019, Chang2019}, redshift-space distortions \cite{Yang2004, Reid2014, Shi2018, Zhai2019} and the abundance and clustering of galaxy groups/clusters \cite{Rykoff2014, Luo2018, Costanzi2019, Bocquet2019, Chiu2020, Luo2020}. 
When a combination of these measurements is employed, it becomes possible to disentangle potential degeneracies among various cosmological parameters.

Among the aforementioned combinations, the most straightforward approach is to combine galaxy clustering and galaxy-galaxy lensing. Galaxy-galaxy auto-correlation and galaxy-mass cross-correlation are defined with respect to the underlying matter-matter correlation function by a factor of $b^2(r)$ and $r_{\rm gm}b(r)$, respectively (where $b(r)$ represents the scale-dependent galaxy bias and $r_{\rm gm}$ is the cross-correlation coefficient). Consequently, this particular combination allows us to break the degeneracy between galaxy bias and the matter-matter auto-correlation function, thus enabling us to determine the cosmological parameters. Recent studies along this line (e.g. \cite{Cacciato2013, Mandelbaum2013, Leauthaud2017, Abbott2018, Lange2019b, Singh2020}) have indicated a discrepancy between the derived cosmological parameters and those inferred from the Planck 2018 cosmic microwave background (CMB) primary anisotropy data \cite{Planck2018}. For instance, the structure growth parameter $S_8=\sigma_8 \sqrt{\Omega_{\rm m}/0.3}$ is consistently found to be approximately $2 - 4\sigma$ lower than the results obtained from Planck \cite{Reid2014, Fu2014, Jee2016, Hildebrandt2017, Hikage2019, Heymans2021}, dubbed ``$S_8$'' tension. Intriguingly, this lower value of $S_8$ is also favored by the pairwise velocity measurements from the 2dF-GRS observations conducted over a decade ago (e.g. \cite{Yang2004}). This tension may suggest the presence of unknown systematic errors in clustering and lensing measurements or the existence of new physics beyond the standard $\Lambda$CDM model \cite{Leauthaud2017}.

For a better assessment of the significance of such tension, it is crucial to utilize the latest observations to improve the precision of $S_8$ constraints. To achieve this, one may need precise measurements of 3D two-point correlation functions (2PCFs) and high signal-to-noise ratio galaxy-galaxy lensing signals. However, the 2PCFs measurements in spectroscopic samples are subject to the restricted survey volume, leading to Poisson noise and cosmic variance issues, especially at higher redshifts.  Alternatively, we recently introduced a novel approach that utilizes two sets of projected 2PCFs to obtain intrinsic clustering for photometric redshift galaxy samples (e.g. \cite{Wang2019}).
This method has been successfully applied to the DESI Legacy Imaging Surveys Data Release 8,  \Authorfootnote providing highly precise projected 2PCFs for multiple galaxy samples defined in bins of color, luminosity, and redshift \cite{Wang2021clustering}.

Additionally, the DESI Legacy Imaging Surveys offer the most extensive collection of galaxy images, which can be utilized to extract highly precise galaxy-galaxy lensing signals. 
Very recently, employing the {\tt Fourier\_Quad} method
\cite{Zhang_2011,Zhang_2015,Zhang2019,Li_2021}, we have obtained the most up-to-date and comprehensive shear catalog from the DESI Legacy Imaging Surveys. The shear estimators of {\tt Fourier\_Quad} are developed based on the multipole moments of the galaxy's 2D power spectrum, allowing for accurate correction of the point spread function (PSF) effect and noise contamination in a model-independent manner \cite{Zhang_2011}. The accuracy of the method has undergone extensive tests using galaxies and PSFs of various forms \cite{Zhang_2015,Li_2021}. 
A direct test has also been performed on the CFHTLenS imaging data \cite{Heymans2012,Erben_2013} using the field-distortion signal \cite{Zhang2019}.

In this study, our aim is to study the dependence of cosmological parameters ($\Omega_m$ and $\sigma_8$) on lens sample redshift through a combined analysis of projected 2PCFs $w_{\rm p}(r_{\rm p})$ and galaxy-galaxy lensing $\Delta\Sigma(\rp)$ measured from DESI Legacy Imaging Surveys Data Release 9. The extensive lens and source sample sizes provide us with the ability to tightly constrain the cosmological parameters for each lens sample. Within the standard $\Lambda$CDM framework, the cosmological parameters are expected to be independent of the properties of the lens and source samples. Any dependence of the cosmological parameters on the properties of the sample can serve as a diagnostic tool for our methodology or as a means to investigate the ``$S_8$ tension".

The structure of this paper is organized as follows. In Section~\ref{sec:data}, we introduce the imaging survey, describe the construction of lens samples, and mitigate the spurious correlation in lens clustering. Section~\ref{sec:shear} covers the shear catalog and lensing measurements. In Section~\ref{sec:method}, we present the joint modeling of clustering and galaxy-galaxy lensing. Our primary findings are presented in Section~\ref{sec:results}. We discuss potential caveats in measurements and modeling in section~\ref{sec:diss}. Finally, we summarize the key conclusions in Section~\ref{sec:summary}. Throughout this work, in addition to the free parameters $\Omega_m$ and $\sigma_8$, we hold other cosmological parameters fixed using the cosmological parameters of Planck 2018 in the following: $n_{\rm s}=0.965$ and $h=H_0/(100 \kmsmpc) = 0.674$. We use $\log$ for the base-10 logarithm.

\begin{figure*}[!t]
\centering
\includegraphics[width=\textwidth]{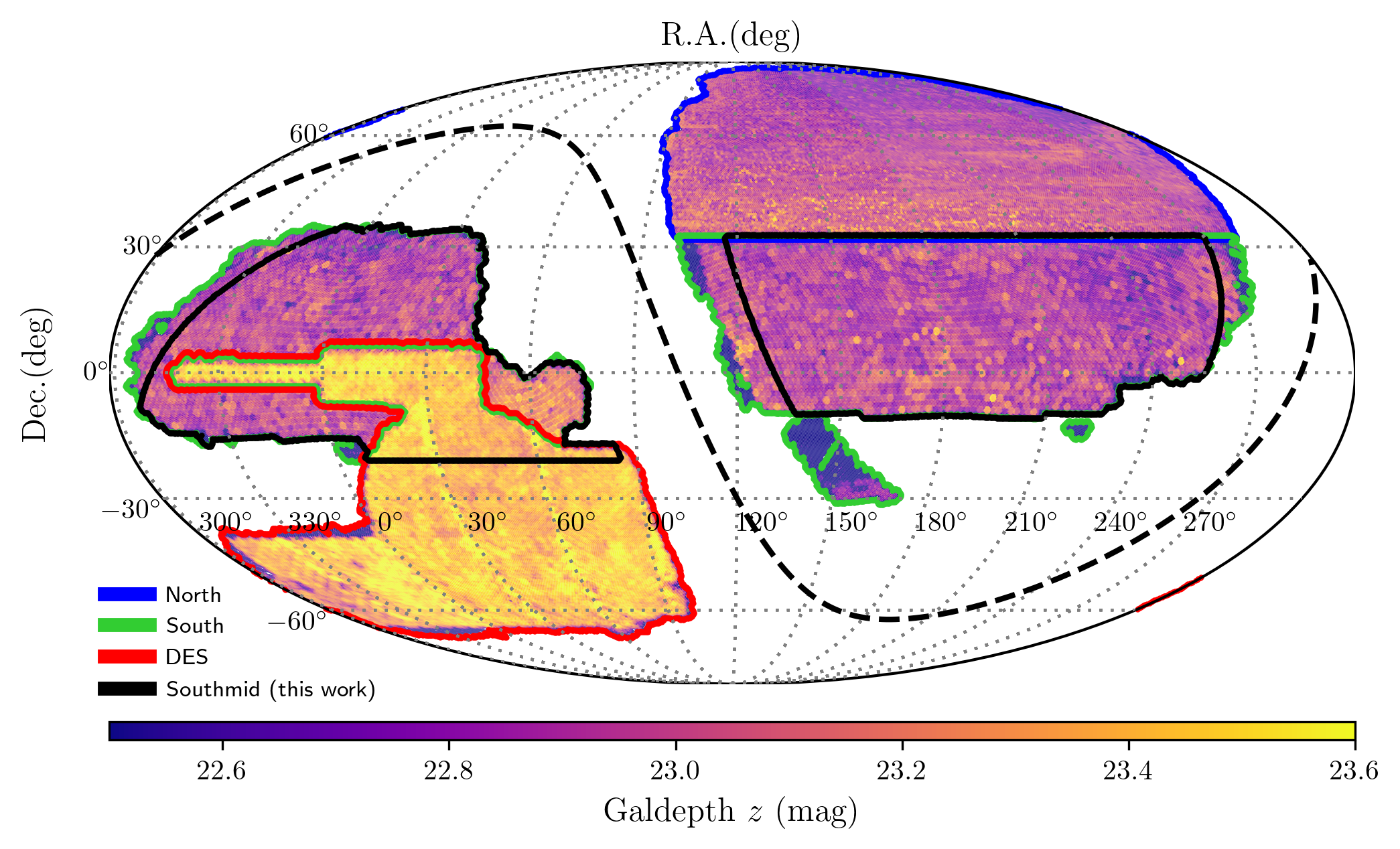}
\caption{The footprint of the Legacy Imaging Survey Data Release 9.
The optical imaging data is from three projects: BASS, MzLS, and DECaLS. 
BASS covers the NGC regions with Dec. $>$ 32.375 degrees (referred to as North, indicated by blue contours) with the
$g$ and $r$ bands. MzLS complements the BASS with $z$ band observation.
DECaLS (green and red contours combined) provided data for the $g$, $r$ and $z$ bands in the NGC regions with Dec. $<$ 32.375 degrees and the entire SGC.
Within DECaLS, the DES region (red contour) has more exposures (four times each band) compared to the non-DES part (referred to as the South, indicated by green contours).  
The black contour represents the footprint of our lens sample, named Southmid (sky coverage
$\sim 9932$ square degrees). The footprint is determined primarily by the distribution of the shear catalog. 
We have color-coded the footprint based on the galaxy depth in the {\it z} band.
}
\label{fig:galdepth_z}
\end{figure*}

\section{Imaging survey, lens sample construction, 
photometric redshift estimation and performance, 
imaging systematics mitigation, and clustering measurements}
\label{sec:data}
In this section, we begin by introducing Data Release 9 of the DESI Legacy Imaging Survey in section~\ref{subsec:dr9}. Moving on to section~\ref{subsec:sample_photoz}, we provide details about the construction of lens samples and a brief overview of the photometric redshift estimation method. In section~\ref{subsec:photoz_err}, we evaluate the performance of photometric redshift (photo-$z$) and analyze the distribution of its errors of our lens samples. Addressing the potential bias in our clustering measurements at large scales, section~\ref{subsec:imaging} focuses on mitigating imaging systematics. Finally, in section~\ref{subsec:2pcf}, we present the measurements of projected 2PCFs and construction of their covariance matrices.

\subsection{Data Release 9 of the DESI Legacy Imaging Surveys}
\label{subsec:dr9}
The DESI Legacy Imaging surveys provide the target catalogs for the 
current going-on DESI surveys. 
Target selection uses a combination of three optical ({\it g/r/z}) bands 
and two mid-infrared ($W1$ 3.4 $\mu$m, $W2$ 4.6 $\mu$m) bands. 
The {\it g/r/z} bands reach a 5$\sigma$ detection of 24/23.4/22.5 AB magnitude
with an exponential light profile of half-light radius $0^{''}.45$.
The optical imaging data comes from three public projects: 

\begin{itemize}
\item the Beijing-Arizona Sky Survey (BASS) observed the North Galactic Cap (NGC) with Dec. $>$ 32.375 degree in {\it g} and {\it r} bands;
\item the Mayall $z$-band Legacy Survey (MzLS) complemented the BASS survey with {\it z} band observation;
\item the DECam Legacy Survey (DECaLS) observed in {\it g/r/z} bands in NGC with Dec. $<$ 32.375 degree and the entire South Galactic Cap (SGC). 
Note that the DECaLS includes the imaging data from the Dark Energy Survey (DES), 
which has more exposure times (thus deeper) than the non-DES part.
\end{itemize}

\begin{table} [H]
\caption{The median values of galaxy depths and PSF sizes in {\it g/r/z} bands of the Legacy Imaging Survey DR9, as a function of sky region indicated in Fig.~\ref{fig:galdepth_z}. 
}\vspace*{1.5mm}
\begin{tabular}{c|ccc|ccc}
\hline
\multirow{2}*{region} & \multicolumn{3}{c|}{galaxy depth (mag)} & \multicolumn{3}{c}{PSF size (arcsec)}\\
\cline{2-7} 
& {\it g} & {\it r} & {\it z} & {\it g} & {\it r} & {\it z}\\ \hline
BASS              & 24.02  & 23.44 & -     & 1.90  & 1.68 &   -    \\ \hline
MzLS              &   -    &   -   & 22.96 &   -    &   -   & 1.24  \\ \hline
South            & 24.38  & 23.82 & 22.93 & 1.90  & 1.39 & 1.32  \\ \hline
DES               & 24.87  & 24.64 & 23.43 & 1.43  & 1.25 & 1.15  \\ \hline
\end{tabular}
\label{table:depth_and_PSF}
\end{table}

Figure~\ref{fig:galdepth_z} shows the footprint of the Legacy Imaging Survey Data Release 9 (DR9) and Table~\ref{table:depth_and_PSF} lists the seeing and depth as a function of optical bands and surveys. 
The Wide-field Infrared Survey Explorer (WISE) satellite provides infrared data for the whole sky. 
See an overview of the Legacy Imaging Surveys in \cite{Dey2019}.

\subsection{Lens sample construction and photo-$z$ estimation}
\label{subsec:sample_photoz}
The galaxy samples used in this work are selected from galaxy catalogs 
using the software package called The Tractor \cite{Lang2016} for source detection 
and photometry measurements in DR9.
Our lens samples were constructed following a selection criterion similar to \cite{Yang2021}. 
Here, we provide a brief summary of the steps involved. 
Firstly, we identify "galaxies" (extended imaging objects) based on the morphological types provided by the Tractor. To ensure a reliable photo-$z$ estimation, we only consider galaxies with at least one exposure in each optical band. We also exclude objects within ${\rm |b| < 25.0^{\circ}}$ (where ${\rm b}$ represents the Galactic latitude) to avoid regions with high stellar density. Additionally, we remove objects whose flux is affected by bright stars, large galaxies, or globular clusters (maskbits 1, 5, 6, 7, 8, 9, 11, 12, 13 \footnote{\url{https://www.legacysurvey.org/dr9/bitmasks/}}). To achieve a similar sky coverage as the shear catalog (see section~\ref{sec:shear}), we further limit our lens sample to -10 degrees $<$ Dec. $<$ 32.375 degrees in the DECaLS-NGC and Dec. $>$ -20 degrees in the South Galactic Cap. Following \cite{Chaussidon2021}, we refer to this region as Southmid, and its footprint is shown in Fig.~\ref{fig:galdepth_z}. The sky coverage of Southmid is approximately 9932 square degrees. We apply identical selections to the publicly available random catalogs\footnote{\url{https://www.legacysurvey.org/dr9/files/\#random-catalogs-randoms}}.

\begin{figure}[H]
\centering
\includegraphics[width=0.48\textwidth]{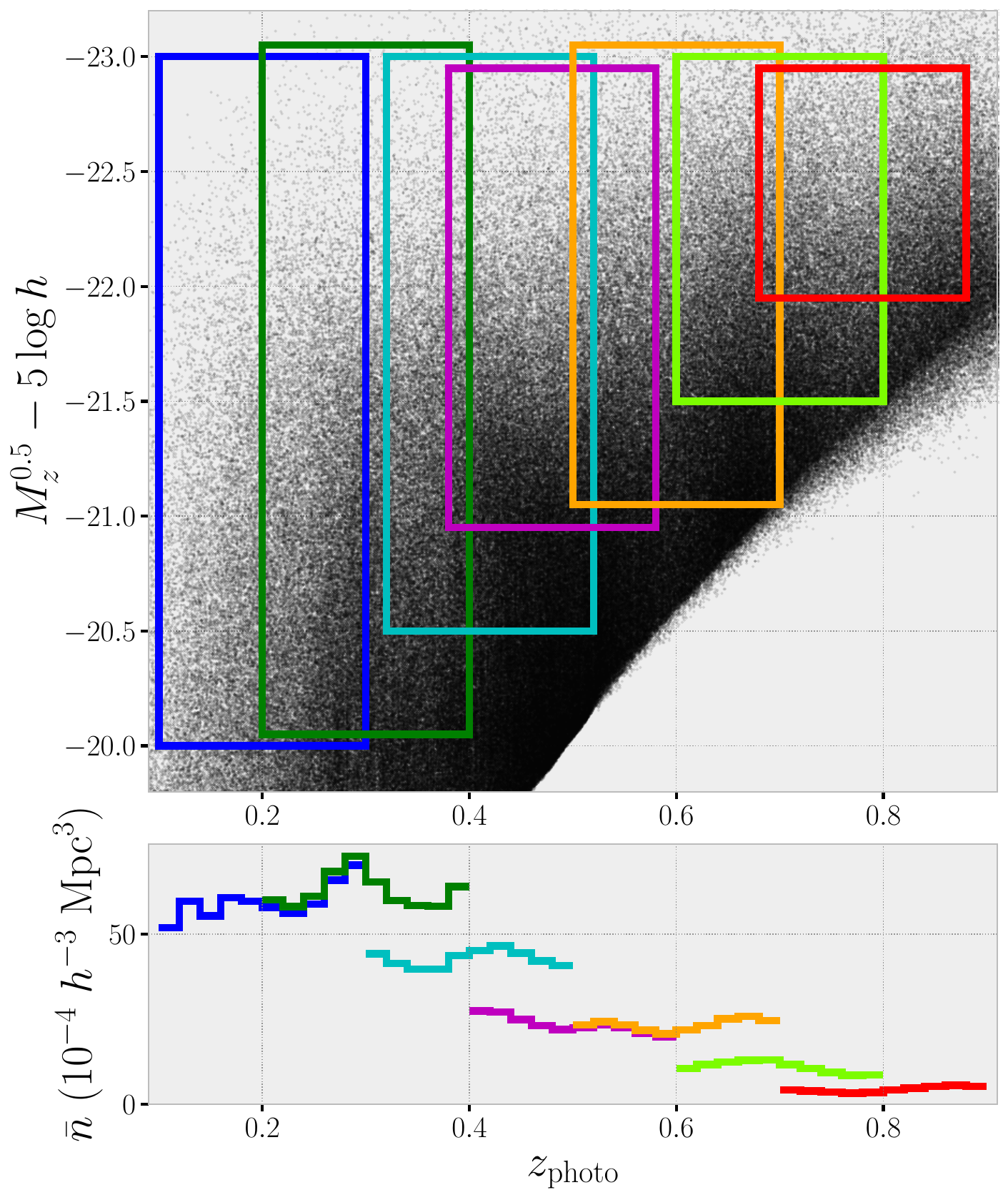}
\caption{
{\it Top}:
Lens sample construction.
To construct volume-limited lens samples in luminosity, we select galaxies 
in bins of photo-$z$ and K-corrected (to $z_{\rm photo}=0.5$) $z$-band absolute magnitude $M_z^{0.5}-5\log h$.
Each patch represents a luminosity-bin volume-limit lens sample, and its
color represents a different redshift range. All samples
have a fixed redshift width of $\Delta z_{\rm photo} = 0.2$.
The maximum luminosity is determined as
$M_z^{0.5}-5\log h = -23$ mag, while the
minimum is determined by the combination of redshift and
the artificial flux limit $z_{\rm mag} < 21$.
We dither the patches for visibility
and the exact lower boundaries in luminosity can be found in Table~\ref{table:sample_info}.
To avoid crowding, the figure includes only 1\% of the luminosity-bin galaxies.
{\it Bottom:} 
We choose red galaxies to form our final lens samples for better photo-$z$ performance. The blue/red galaxy separation is detailed in Appendix~\ref{subsec:appendix1}.
The solid lines show the redshift dependence of the comoving number density of galaxies in each of our lens samples, assuming Planck 2018 cosmology. 
}
\label{fig:vol_limit}
\end{figure}

The lens sample utilizes photo-$z$ provided by the publicly available catalog
Photometric Redshifts estimation for the Legacy Surveys (PRLS \footnote{\url{https://www.legacysurvey.org/dr9/files/\#photometric-redshift-sweeps}}, \cite{Zhou2021}). Photo-$z$ values are obtained via random forest, a machine learning algorithm that learns the mapping of galaxy properties (such as colors, magnitudes, and morphology) to their redshifts. \cite{Zhou2021} established this mapping relationship using eight properties\footnote{They are $r$-band magnitude, $g-r$, $r-z$, $z-W1$, and $W1-W2$ colors, half-light radius, axis ratio, and shape probability.} and secure redshifts available from various spectroscopic surveys and the COSMOS. The total number of galaxies with accurate redshifts cross-matched with the PRLS sources is approximately 1.5 million. However, to avoid bias in the photo-$z$ estimation caused by non-uniform distribution in the multi-dimensional color-magnitude space, only around 0.67 million of them are used. For details of photo-$z$ estimation, we refer the interested reader to \cite{Zhou2021}. The reliability of the photo-$z$ in the PRLS catalog has been well tested for luminous red galaxy (LRG) samples. Furthermore, they demonstrated reasonably good photo-$z$ performance for low redshift galaxies with $z_{\rm mag}<21$, achieving a $\sigma_{\rm NMAD}$ of approximately 0.013 and an outlier rate of about 1.51\% \footnote{The $\sigma_{\rm NMAD}$ is defined as 
$\sigma_{\rm NMAD} = 1.48\times{\rm median}(|z_{\rm photo}-z_{\rm spec}|)/(1+z_{\rm spec})$. 
The outlier rate is defined $|z_{\rm photo}-z_{\rm spec}| > 0.1\times(1+z_{\rm spec})$. Note that the definition of an outlier rate is different in the latter part of this work.} (refer to their Figures B1 and B2). To construct the lens sample with reliable photo-$z$, we select galaxies with $z_{\rm mag}<21$ and $z_{\rm photo} < 0.9$. 
It is worth noting that the default $z_{\rm photo}$ used in this work are the
median values of the photo-$z$ estimation in the PRLS catalog.
Additionally, if secure redshifts are available, we replace photo-$z$ with
secure redshifts.

On top of the magnitude and redshift selection, we specifically choose red galaxies for our analysis for better photo-$z$ performance. 
To accomplish this, we utilize $z$-band absolute magnitude $M^{0.5}_z-5\log h$ (K-corrected to $z = 0.5$ by \cite{Blanton2007}) and the
photo-$z$. By combining these factors, we create seven volume-limited lens samples, as illustrated in the upper panel of Fig.~\ref{fig:vol_limit}.
These samples range from $z=0.1$ to $z=0.9$, with a consistent bin width of $\Delta z_{\rm photo} = 0.2$.
Given that photo-$z$ data from \cite{Zhou2021} are more reliable for LRGs, we specifically select red galaxies from these luminosity-bin samples. By examining the color-magnitude diagrams for each luminosity-bin sample (Appendix~\ref{subsec:appendix1}), we identify two distinct populations: a loosely distributed blue cloud and a tight red sequence (refer to Fig~\ref{fig:CMD}). We drop the blue galaxies and keep red galaxies, which form the final lens samples.

The lower panel of Fig.~\ref{fig:vol_limit} shows the number density as a function of the redshift for seven lens samples. The number density of most samples have little dependence on redshift, except for the first two samples. For simplicity, we ignore the redshift evolution of the clustering measurements within the redshift bin of each lens sample, as is usually done in a few previous 2$\times$2pt or 3$\times$2pt analysis e.g. \cite{Mandelbaum2013, Sugiyama2022, Miyatake2021, Heymans2021, Porredon2021}.

In summary, our lens samples consist of luminous and red galaxies, and their redshift performance is expected to be comparable to that of LRGs. For further details, refer to Table~\ref{table:sample_info}, which provides information on the lens samples.

\subsection{Lens photo-$z$ performance and photo-$z$ error distributions}
\label{subsec:photoz_err}
\begin{figure}[H]
\centering
\includegraphics[width=0.48\textwidth]{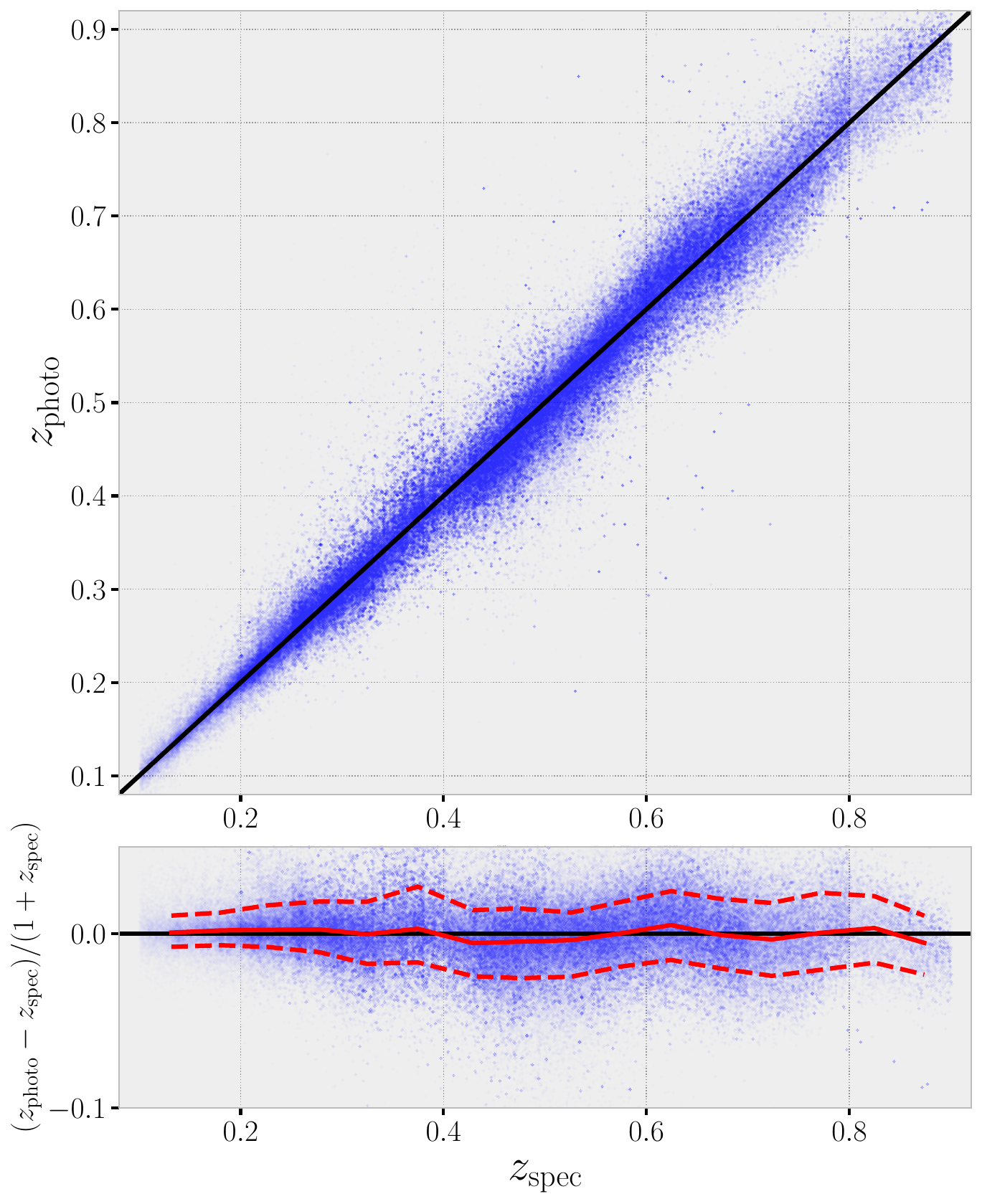}
\caption{
{\it Top}: photometric redshift VS. spectroscopic redshift of weighted spectroscopic 
subsample of all lens samples combined.
The spectroscopic subsample is weighted such that it matches with photometric sample in multi-dimensional color-magnitude space. 
Photo-$z$ performance evaluated from a weighted spectroscopic subsample is believed to represent the photometric sample.
{\it Bottom}: The normalized redshift difference, 
$(z_{\rm photo}-z_{\rm spec})/(1+z_{\rm spec})$, as a 
function of spectroscopic redshift $z_{\rm spec}$. 
The solid (dashed) red line is the median (68 percentile).
The $\sigma_{\rm NMAD}$ and outlier rate of the weighted 
subsample are $\sim 0.017$ and $\sim 0.7$\%, respectively.
}
\label{fig:lens_photoz_performance}
\end{figure}

We evaluate the photo-$z$ performance of lens galaxies by comparing their 
photo-$z$ and spectroscopic redshift (spec-$z$) values. However, it is important to note that spectroscopic subsamples tend to consist of brighter objects compared to the entire photometric sample. Consequently, a direct comparison might not accurately represent the entire lens sample. To address this issue, we apply weights to the spectroscopic galaxies to ensure that the weighted spectroscopic subsamples match with the photometric sample in multi-dimensional color-magnitude space \cite{Lima2008, Bonnett2016}. These weights are assigned following the methodology presented in \cite{Zhou2021}. 
We summarize as follows.
For each photometric galaxy, we link it to its nearest spectroscopic neighbor based on their similarities in color-magnitude space of $r$-band magnitude, $g-r$, $r-z$, $z-W1$, and $W1-W2$. The weights assigned to spectroscopic galaxies are determined by the number of photometric galaxies they are linked to. The photo-$z$ performance of the weighted spectroscopic subsamples is depicted in Fig.~\ref{fig:lens_photoz_performance}. In general, photo-$z$ performance demonstrates satisfactory results, with $\sigma_{\rm NMAD} \sim 0.017$ and an outlier rate of approximately 0.7\%.

\begin{figure*}[!t]
\centering
\includegraphics[width=\textwidth]{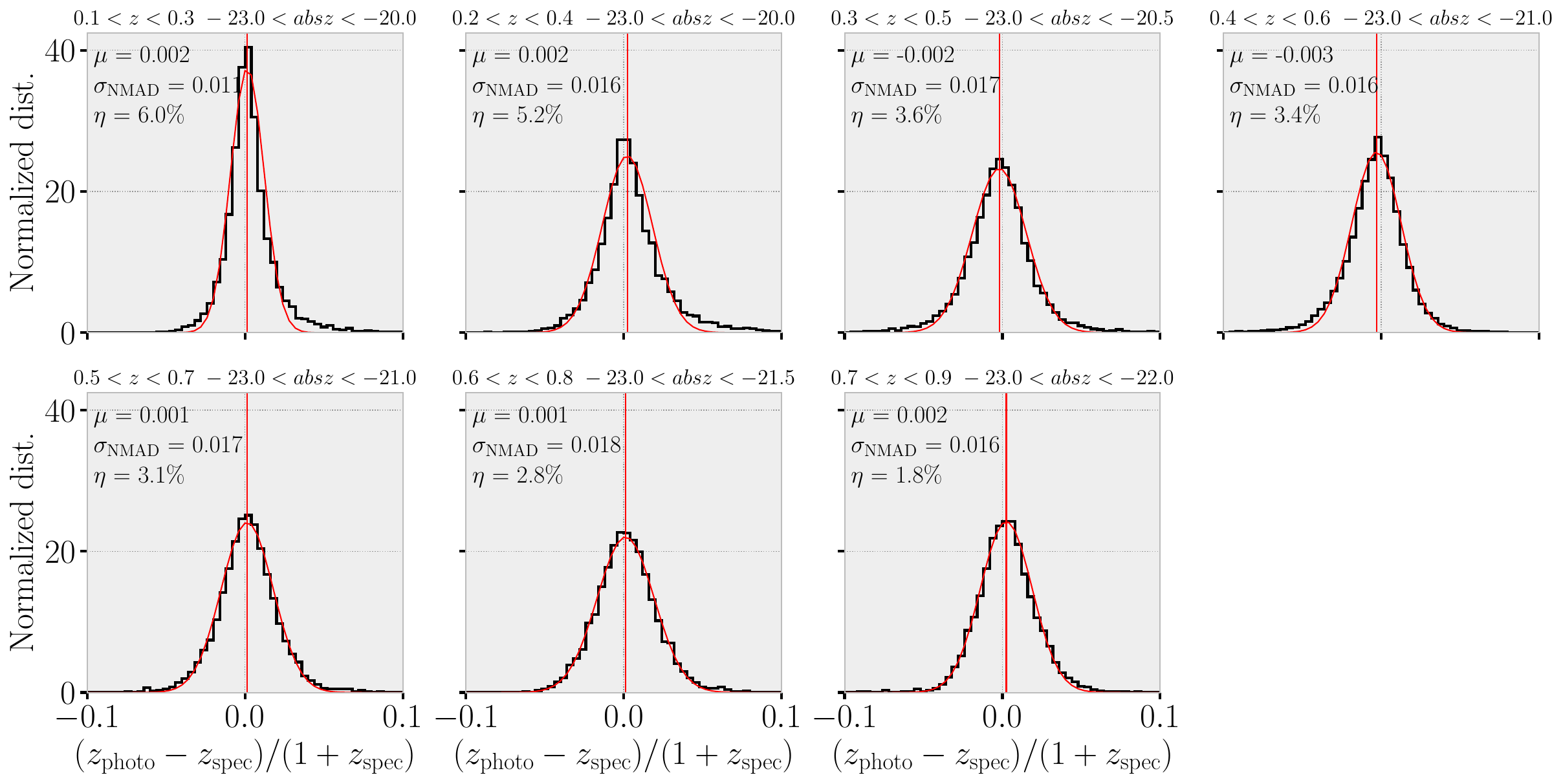}
\caption{
The distributions of normalized redshift difference (black histograms), 
$(z_{\rm photo}-z_{\rm spec})/(1+z_{\rm spec})$, of seven lens samples.
The information of lens samples are indicated on top
of each panel, where $z$ stands for $z_{\rm photo}$ and absz for 
K-corrected (to $z = 0.5$) $z$-band absolute magnitude $M^{0.5}_z-5\log h$.
In the upper left of each panel, $\mu$ (vertical line) and $\sigma_{\rm NMAD}$ (1-$\sigma$ width of the red curve line)  are the median and normalized median absolute deviation of histograms, respectively. 
The red curve is the corresponding Gaussian distribution with mean $\mu$
and scatter $\sigma_{\rm NMAD}$. 
Note that the red curve is not a fit to the histograms.
Therefore, the resemblance in shape between the histograms and the red curves suggests that
the normalized redshift difference closely follows a Gaussian distribution.
A nonzero $\mu$ indicates a bias in the normalized redshift difference, while
$\sigma_{\rm NMAD}$ characterizes the effective uncertainty of photo-$z$, 
which can be compared with constraints derived from clustering measurements 
(e.g. Fig.~\ref{fig:sigmaz_compare}). 
Furthermore, we calculate the rate of catastrophic outliers $\eta$ for each sample, defined as $|z_{\rm photo}-z_{\rm spec}|/(1+z_{\rm spec}) > 3\sigma_{\rm NMAD}$. 
The presence of catastrophic outliers may lead to an underestimation of intrinsic clustering strength (for instance, correlated galaxy pairs go outside of integration depth).
Assuming that these outliers randomly distribute in the redshift direction, we then scale the model prediction by a factor of $(1-\eta)^2$ to account for such an underestimation (refer to Section~\ref{subsec:wp_model} for detailed information).
When plotting, we use weighted
spectroscopic subsamples that match each lens sample in the color-magnitude space.
}
\label{fig:photoz_err}
\end{figure*}

In modeling the impact of photo-$z$ on clustering (see details in section~\ref{subsec:wp_model}), we make the assumption that the photo-$z$ values of our lens galaxies follow a Gaussian distribution with respect to their true redshift, described by the equation:
\begin{equation}
P\left(z_{\rm photo} - z_{\rm spec}\right)=\mathcal{N} \left(0, \sigma_z (1 + z_{\rm spec}) \right)
\label{eq:GaussianPhotoz}
\end{equation}
Here, $\mathcal{N}(\mu, \sigma)$ represents a Gaussian distribution with a mean of $\mu$ and a scatter of $\sigma$. The parameter $\sigma_z$ represents the effective photo-$z$ uncertainty of the lens sample. To assess the validity of the Gaussian assumption, we plot the distributions of the normalized redshift difference, $(z_{\rm photo}-z_{\rm spec})/(1+z_{\rm spec})$, as shown in Fig.~\ref{fig:photoz_err}. We utilize weighted spectroscopic subsamples for plotting.

In general, the distributions closely resemble a Gaussian distribution, with the exception of a minor bias on the order of $\sim$ 0.001 present in all samples. The width of the distributions, characterized by $\sigma_{\rm NMAD}$, encodes
the photo-$z$ uncertainty from the color-magnitude-redshift relation, which can then be compared with the photo-$z$ error $\sigma_z$ constrained from the large-scale structure (Section~\ref{subsec:wp_model}). The comparison between these two quantities is presented in Section~\ref{subsec:model_test}.

\begin{figure*}[!t]
\centering
\includegraphics[width=\textwidth]{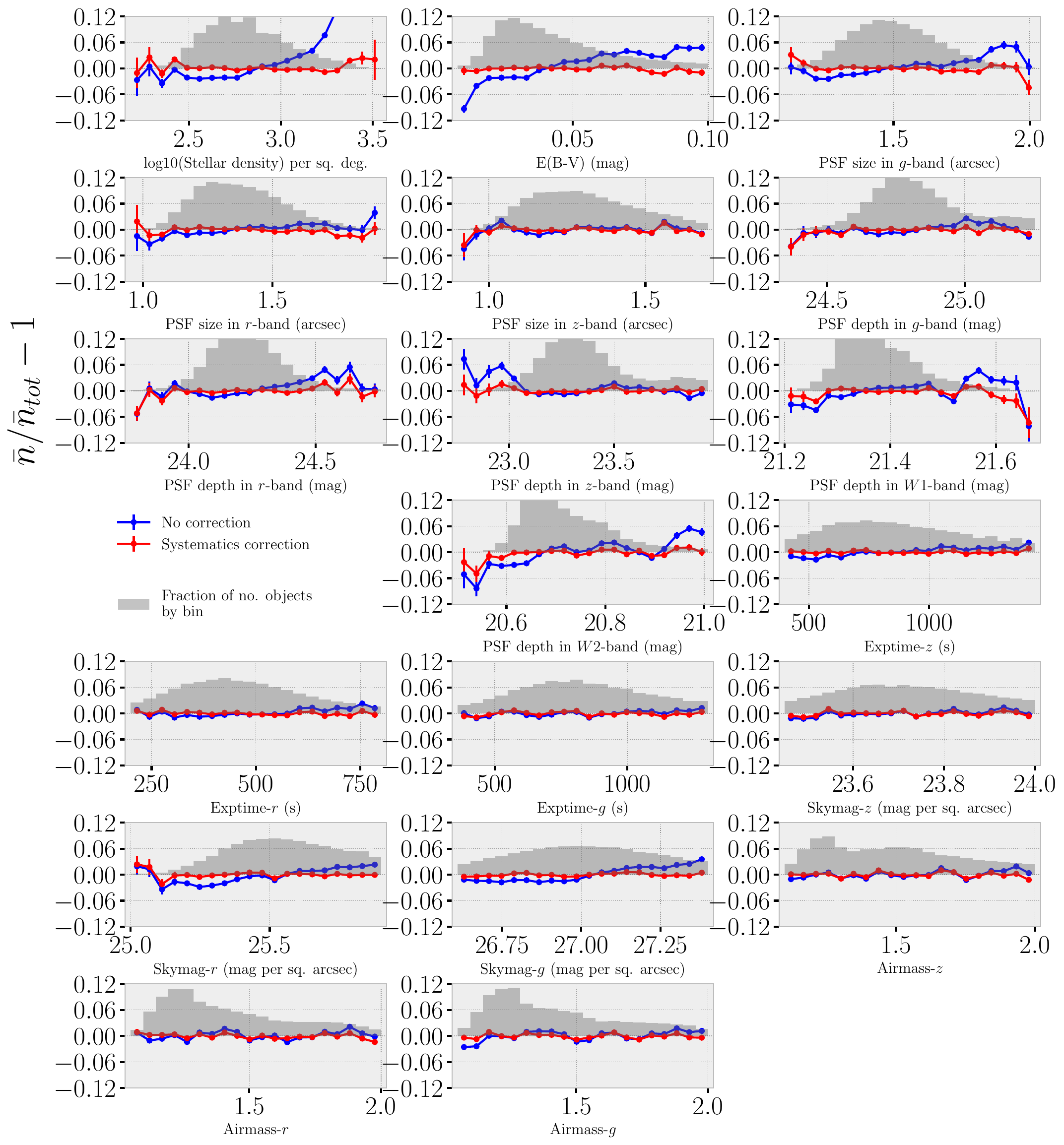}
\caption{
Illustration of imaging systematics mitigation for the lens sample $0.5 < z_{\rm photo} < 0.7$. 
The blue and red lines show the surface density
relative to global mean, before and after the systematics correction,
in bins of 19 input imaging maps. 
The grey histogram is the fraction of valid pixels (or galaxies) in bins of imaging maps, 
which serves to estimate the error bars with the standard deviation.
After mitigation, the corrected samples show little fluctuation with respect to the
imaging properties.
Since imaging systematics mitigation varies with sample's properties,
we individually apply imaging systematics mitigation procedure
to each lens samples before measuring the correlations. 
Other lens samples show similar results.
}
\label{fig:imaging_systematics}
\end{figure*}

\subsection{Imaging systematics mitigation}
\label{subsec:imaging}

Observational conditions, such as stellar contamination, Galactic foreground, and seeing, are known to introduce spurious fluctuations in the observed galaxy densities (e.g. \cite{Scranton2002a, Myers2006, Ho2012, Morrison2015, Ross2011a, Ross2017}). These spurious fluctuations significantly contaminate the clustering measurements at large scales, thereby biasing the inference of cosmological parameters.

One approach to mitigate imaging systematics is to assign weights to galaxies in photometric samples, ensuring that the weighted samples exhibit minimal dependence on imaging properties. Multi-linear (or -quadratic) regression is commonly adopted to model the dependence of observed galaxy surface densities on various imaging maps \cite{Myers2015, Prakash2016, Reid2016, Ross2020}. The weights can be obtained by minimizing the difference between observed densities and a multi-linear (or -quadratic) function of imaging maps.

However, a linear (or quadratic) model may fail to capture higher-order dependencies on imaging features in strongly contaminated regions, such as those near the Galactic plane (see, e.g. \cite{Ho2012}). Recent studies have begun to employ machine learning algorithms to capture the complex relationship between observed densities and imaging maps.

In our case, we employ the Random Forest (RF) mitigation method developed by \cite{Chaussidon2021} to our lens samples for two reasons: First, RF outperforms linear or quadratic models; second, compared to the Neural Network approach, RF achieves similar mitigation results at a significantly lower computational expense. For a detailed comparison, we refer interested readers to \cite{Chaussidon2021}. We provide a brief overview of the main steps here, while a comprehensive description of the methodology is available in \cite{Chaussidon2021}. Furthermore, it should be noted that most of the following procedures are conveniently encapsulated in the publicly available code {\tt regressis}\footnote{\url{https://github.com/echaussidon/regressis}}.

\begin{figure*}[!t]
\centering
\includegraphics[width=\textwidth]{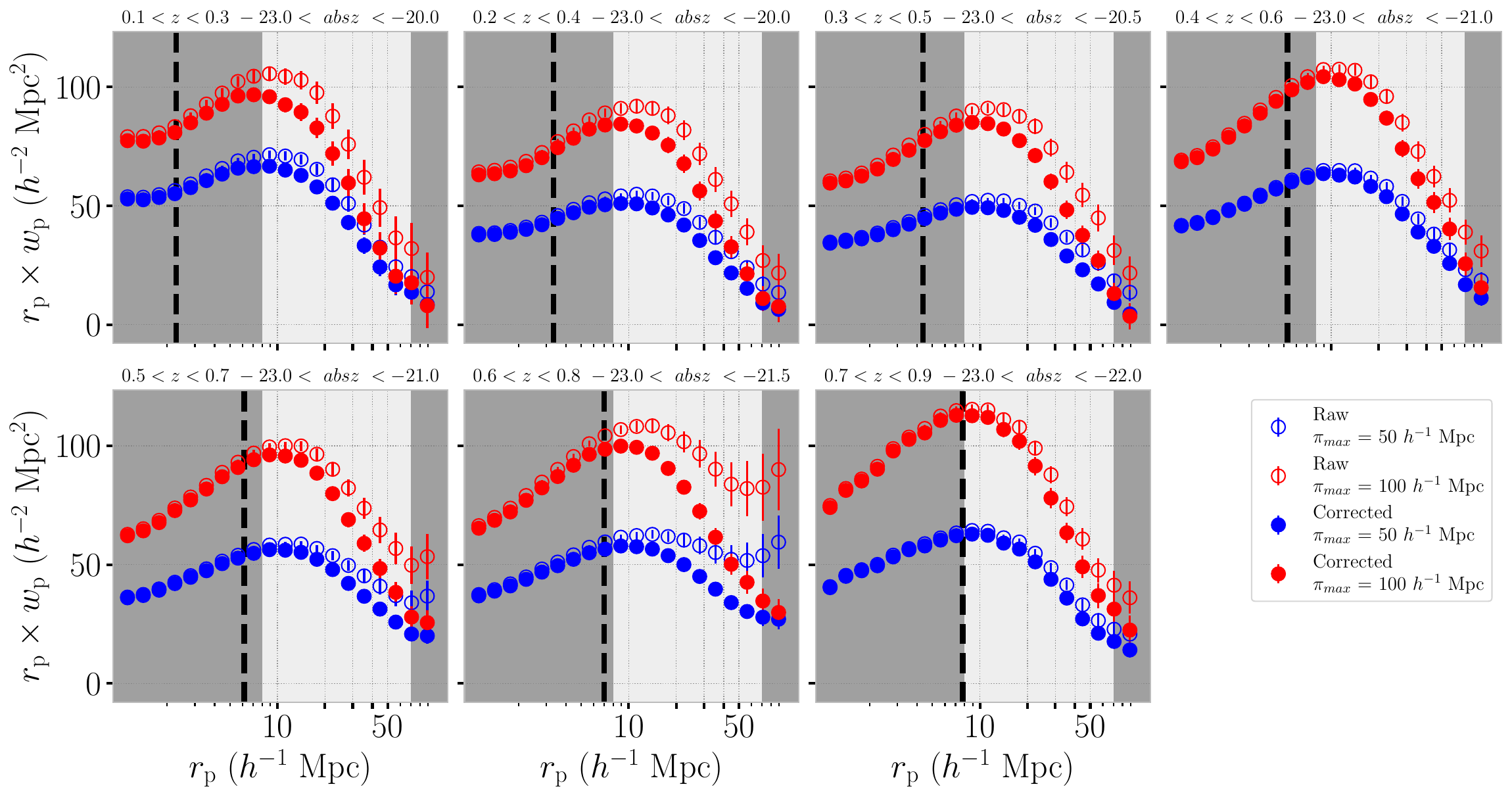}
\caption{
The measured projected 2PCFs of lens samples, before (empty circles) and after (filled circles) imaging systematic correction. The blue and red symbols represent the projected 2PCFs integrated to $\pi_{\rm max} = 50~h^{-1}{\rm Mpc}$ and $\pi_{\rm max} = 100 ~h^{-1}{\rm Mpc}$, respectively. 
The correction is quite substantial at large scales. The vertical dashed line marks the scale
(pixel size $\sim 0.23$ deg) above which the spurious correlation introduced by imaging systematics is largely mitigated. The light-shaded region represents the range of separations used for the estimation of the cosmological parameters: $r_{\rm p} = [8, 70]~h^{-1}{\rm Mpc}$. The errors are estimated using the jackknife method with $N_{\rm jkf} = 200$. 
}
\label{fig:wp_w_sys_weight}
\end{figure*}

\begin{itemize}
\item (1) The imaging maps utilized in this study are kindly provided by 
E. Chaussidon (private communication). These maps were generated
from the code script \texttt{bin/make\_imaging\_weight\_map} from the {\tt desitarget}\footnote{\url{https://github.com/desihub/desitarget}} package with {\tt HEALPix}\cite{Healpix} $N_{\rm side}=256$ (resolution $\sim 0.23$ deg). We have selected 19 imaging maps, which are as follows:
stellar density \cite{Gaia2018}, Galactic extinction\cite{Schlegel1998, Schlafly2011}, sky brightness (in $g/r/z$ band), airmass ($g/r/z$ band), 
exposure time ($g/r/z$ band), PSF size ($g/r/z$ band) and PSF depth ($g/r/z$/W1/W2 band). The lens samples are binned at the same resolution.

\item (2) The total number of pixels within the Southmid is 187211 (black contour in Fig.~\ref{fig:galdepth_z}).
Among these, we only consider pixels with $f_{{\rm pix}, i} > 0.8$,
where $f_{{\rm pix}, i}$ is the fractional observed area in pixel $i$.
This value is calculated as the ratio of random points after and before maskbit selection (Section~\ref{subsec:sample_photoz}).
This selection ensures that the regression is limited to reliable pixels. Furthermore, we exclude pixels with values of {\tt NAN} on any of the imaging maps. In total, we remove 8219 pixels, accounting for 4.39 percent of the lens footprint.

\item (3) To prevent overfitting, we employ the $K$-fold cross-validation technique. Specifically, we adopt $K = 14$.
This choice of folds ensures efficient regression analysis while reducing the risk of overfitting. Given that imaging systematics vary across locations, we ensure that each fold contains pixels from across the entire footprint.

\item(4) We utilize the same RF hyperparameters as in \cite{Chaussidon2021}, employing 200 decision trees with a minimum of 20 pixels at a leaf node. We have experimented with different numbers but found no significant differences in imaging systematics reduction.

\end{itemize}

In valid pixels, {\tt regressis} returns the weights to correct the imaging systematics.
However, due to differences in redshift and luminosity, samples may experience varying levels of imaging contamination. Therefore, we apply the mitigation procedures individually to each lens sample. In Fig.~\ref{fig:imaging_systematics}, we present the mitigation results for one of the seven lens samples with a redshift range of $0.5 < z_{\rm photo} < 0.7$. Prior to mitigation, the mean galaxy density, categorized by imaging properties, exhibits a systematic trend (indicated by the blue lines). However, after mitigation, this trend is substantially reduced (represented by the red lines). To ensure the reliability of our results, we performed a sanity test and confirmed that the cross-correlation between the corrected samples and all imaging maps is close to zero.

\subsection{Clustering measurements in the lens samples}
\label{subsec:2pcf}

To minimize the impact of redshift space distortion (RSD),
we calculate the projected 2PCFs $w_{\rm p}^{\rm obs}(r_{\rm p}|\pi_{\max})$ 
by integrating the redshift-space 2PCFs $\xi^{\rm obs}(r_{\rm p},\pi)$:
\begin{equation}
\label{eq:wp}
w_{\rm p}^{\rm obs}(r_{\rm p}|\pi_{\max}) = 2\int_0^{\pi_{\max}}\xi^{\rm obs}(r_{\rm p},\pi){\rm d}{\pi}.
\end{equation}
where $r_{\rm p}$ and $\pi$ represent the transverse and line-of-sight separations between galaxy pairs, respectively. $\pi_{\max}$ is the upper bound of the integration length along the line-of-sight direction. 
Apart from RSD, the measured projected 2PCFs are affected by inaccurate photo-$z$, which can significantly underestimate the clustering signal. To account for the photo-$z$ effect on clustering, we adopt a novel method developed in \cite{Wang2019, Wang2021clustering}, which involves two sets of projected 2PCFs with different integration limits $\pi_{\max}$. 
See the detailed formalism in Section~\ref{subsec:wp_model}.
In this study, we choose $\pi_{\max}=50$ and $100~h^{-1}\rm Mpc$.

The redshift-space 2PCFs is given by the Landy-Szalay estimator \cite{Landy93}:
\begin{equation}
\xi^{\rm obs}(r_{\rm p},\pi) =\frac{a{\rm DD}(r_{\rm p},\pi)-2b{\rm DR}(r_{\rm p},\pi)+{\rm RR}(r_{\rm p},\pi)}{{\rm RR}(r_{\rm p},\pi)},
\end{equation}
where DD, DR, and RR represents the weighted pairs of 
galaxy-galaxy, galaxy-random, random-random in a given $(r_{\rm p},\pi)$ pair separation bin. 
Normalization terms are defined as follows:
\begin{equation}
a=\frac{\sum\limits_{i \neq j} w_{i}^{R} w_{j}^{R}}{\sum\limits_{i \neq j} w_{i}^{D} w_{j}^{D}} \quad \text{and} \quad b=\frac{\sum\limits_{i \neq j} w_{i}^{R} w_{j}^{R}}{\sum\limits_{i} w_{i}^{D} \sum\limits_{j} w_{j}^{R}},
\label{eq:LS}
\end{equation}
where $w_i^{D}$ is the imaging correction weight for pixel $i$ in the data and,  $w_i^{R} \equiv 1$ for random.
Note that we have applied the identical footprint cuts, number of exposure times
restrictions, and maskbits to the random catalogs as in lens sample construction (cf. Section~\ref{subsec:sample_photoz}).
The number of random galaxies is five times that of the galaxies.
We use Python package \texttt{Corrfunc} \cite{Sinha2020} for counting pairs.
We set linearly spaced bins in $\pi$ with a bin width of $\Delta \pi = 1~h^{-1}\rm Mpc$, ranging from 0 to $100~h^{-1}\rm Mpc$, and 20 logarithmically spaced $r_{\rm p}$ bins from 1 to $100~h^{-1}\rm Mpc$.

Figure~\ref{fig:wp_w_sys_weight} illustrates the comparisons between the measured projected 2PCFs before and after mitigation of imaging systematics. 
The correction is negligible on small scales but substantial on large scales, exceeding 50\% in some cases.
To estimate the cosmological parameters, we use measurements within the scales $r_{\rm p} = [8, 70]~h^{-1}{\rm Mpc}$. These scale cuts are motivated by the ``minimal-bias" model \cite{Sugiyama2020}, which successfully recovered cosmological parameters ($\sigma_8$ and $\Omega_m$) within a 68\% credible interval by employing the linear galaxy bias and the non-linear matter power spectrum. Further discussion is provided in section~\ref{sec:method}. It is important to note that imaging contamination is mostly mitigated above the pixel scale of approximately 0.23 degrees (corresponding comoving scales marked as vertical dashed lines in Fig.~\ref{fig:wp_w_sys_weight}). The chosen minimal scale cuts for each lens sample are safely above this threshold.

To estimate the covariance matrix, we employ the jackknife resampling method. The Southmid footprint is divided into $N_{\rm jkf}$ = 200 spatially contiguous and equally sized sub-regions. We measure the projected 2PCFs 200 times, leaving out one different sub-region each time, and calculate the covariance as 199 times the variance of the 200 measurements.

\section{Shear catalog and galaxy-galaxy lensing measurements}
\label{sec:shear}
\subsection{Shape measurements and shear catalog}
The shear catalog used in this work is measured from DECaLS\footnote{
\url{https://www.legacysurvey.org}. Images were observed between 2013 and 2019.} 
images using the {\tt Fourier\_Quad} pipeline \cite{Zhang2022_decals}. 
The source positions are provided by \cite{Zou2019}.
To comply with the clustering measurement, we used photo-$z$ estimates from the PRLS catalog, described in Section~\ref{subsec:sample_photoz}. The source catalog in \cite{Zou2019} applied a $r$-band magnitude cut of $r_{\rm mag} < 23$ for source detection and photometric estimation. The photo-$z$ quality cut of $z_{\rm mag} < 21$ is also applied. The {\tt Fourier\_Quad} pipeline was independently applied to the $g$, $r$, and $z$ band imaging data to generate the corresponding shear catalog, covering approximately 10000 square degrees of the sky, as shown in the black contour of Fig.~\ref{fig:galdepth_z}. The exposures of each field were processed independently, eliminating image stacking-related systematics and yielding a source catalog with more sources.

The field distortion, caused by the projection from the celestial sphere to the CCD plane,
manifests itself as a CCD coordinate-dependent shear imposed on the galaxy images. 
This distortion signal can be recovered by averaging among galaxies from all observation fields within the same CCD region. 
\cite{Zhang2019} proposed this as a systematic test for shear measurement, which validated the precision of our shape measurement. 
The field distortion test presented in \cite{Zhang2022_decals}
demonstrated that our shear catalog can recover the distortion signal at a level of $1\times10^{-3}$ with a multiplicative bias of $10^2m = (5\pm3,3\pm3)$ and an additive bias of $10^5c = (0\pm3, -10\pm3)$. 
The presence of a multiplicative bias would introduce a few percent bias in the Excess Surface Density (ESD) measurements, approximately on the order of $(1+m)$.
In this study, we did not consider a correction, since the value is consistent with zero within a 2$\sigma$ level.
A small additive bias was identified, but will be nullified in the tangential coordinate.
Before ESD measurement, the field distortion signal was subtracted.
To address potential additive bias in the ESD measurement, 
we subtracted the ESD measured from a random catalog with a number five times that of the lens galaxies.

In this study, we exclusively utilize the $z$ band source catalog, 
motivated by its data quality and the field distortion test shown in \cite{Zhang2022_decals}. 
To ensure reliable shape measurements, 
we apply the following selection criteria to the galaxies: $z_{\rm{mag}}<21$ and signal-to-noise ratio $\nu_F >4$.
The magnitude cut is to exclude faint galaxies with less reliable shape measurements and photo-$z$ values, while the signal-to-noise ratio cut is to prevent selection bias in the {\tt Fourier\_Quad} method (e.g. \cite{Li2021_snr}).
We also remove the bias induced by image geometry boundaries through the algorithm proposed by \cite{Wang2021_geo}. 
After these selection, the multiplicative bias for each source sample is listed in Table~\ref{table:sample_info}.

To mitigate the potential ESD dilution caused by incorrect photo-$z$ and minimize the contamination of intrinsic alignment, we adopt a relatively large source-lens cut in redshift, i.e. $z_s > z_l + 0.25$. 
Finally, we end up with a sample of 118984459 source galaxies for the ESD measurements. The corresponding shape noise ranges from approximately 0.26 to 0.31, depending on the average size of the point spread function \cite{Zhang2022_decals}.

\subsection{Galaxy-Galaxy lensing measured by {\tt Fourier\_Quad} method}

{\tt Fourier\_Quad} is a moment-based method that measures galaxy shapes by analyzing its power spectrum. 
The method provides five shear estimators: $G_1$, $G_2$, $N$, $U$, $V$ (for more details, please refer to \cite{Zhang2017_pdfsym}).
The main shape estimator, $G_1$ and $G_2$ ($G_{1/2}$ hereafter), encompasses both the intrinsic shape information, 
$G_{1/2}^I$, and the shear-induced part, $g_{1/2}\cdot(N \pm U)$.
Assuming isotropic orientations of galaxies, 
the ensemble average of the intrinsic shape estimator $G_{1/2}^I$ cancels out, resulting in a symmetrical probability distribution function (PDF) centered around zero.
However, the shear signal introduces an anisotropic part, $g_{1/2}\cdot(N \pm U)$,
to the PDF, causing a shift in its symmetry axis, as illustrated
in Figures 3 and 4 of \cite{Zhang2017_pdfsym}.

The PDF\_SYM approach, unbiased and optimal in statistics, 
effectively recovers the signal by symmetrizing the PDF of $G_{1/2}$ \cite{Zhang2017_pdfsym}.
When the anisotropic component in the shear estimators is corrected by the true signal, 
denoted as $\hat{g}_{1/2}=g_{1/2}$, 
the PDF of
\begin{equation}
    \hat{G}_{1/2} = G_{1/2} - \hat{g}_{1/2}\cdot(N \pm U)
\end{equation}
regains its symmetry.
In this process, the asymmetry of the PDF can be quantified using the formula
\begin{equation}
    \chi^2 = \frac{1}{2}\sum_{i=1}^{N}\frac{(n_i - n_{-i})^2}{n_i + n_{-i}},
\end{equation}
where $n_i$ represents the galaxy counts of the $i$-th bin
\footnote{The bins should be symmetrical on both sides of zeros for either $G_{1/2}$ or $\hat{G}_{1/2}$}. 
Therefore, as $\chi^2$ approaches its minimum, the shear signal can be readily determined.

In the context of galaxy-galaxy lensing measurements,
the tangential shear $\gamma_t$ can be derived by symmetrizing the PDF of
\begin{equation}
\hat{G}_{t}(z_s) = G_{t}(z_s) - \hat{\gamma_t}(z_s)\cdot \left( N(z_s) + U_t(z_s) \right),
\label{eq:gamma_t}
\end{equation}
where the tangential shear estimators are defined as,
\begin{align}
    G_t + iG_{\times} &= (G_1 + iG_2)\exp[2i\theta], \\ \nonumber
    U_t + iV_{\times} &= (U + iV)\exp[4i\theta].
\end{align}
Here, $\theta$ denotes the position angle measured from the north pole in the north-east direction.
The tangential shear $\gamma_t$ is connected to the ESD $\Delta\Sigma(\rp)$ 
through the equation:
\begin{equation}
\label{eq:esd_measure}
\Delta\Sigma(\rp) =\gamma_t(z_s)\Sigma_c(z_l,z_s),
\end{equation}
where $\Sigma_c(z_l,z_s)$ represents the critical surface density defined in comoving coordinates as:
\begin{equation}\label{eq:Sigma_c}
\Sigma_c(z_l,z_s)=\frac{c^2}{4\pi G}\frac{\chi_s}{\chi_l\chi_{ls} (1+z_l) }, 
\end{equation}
where $z_l$ and $z_s$ indicates the redshifts of lens and source, respectively. 
$\chi_l$, $\chi_s$ and $\chi_{ls}$ corresponding to the 
comoving angular diameter distance (comoving transverse distance) of the lens, the source, and the distance between the lens and the source, respectively. 
The factor of $1+z_l$ in the denominator appears because the ESD is measured in the comoving coordinates.
By combining Eq.~\ref{eq:gamma_t} and Eq.~\ref{eq:Sigma_c}, the ESD can be restored by symmetrizing the PDF of
\begin{equation}
\hat{G_t}(z_s) = G_t(z_s)-\frac{\Delta\hat{\Sigma}}{\Sigma_c(z_l,z_s)}\cdot \left( N(z_s)+U_t(z_s)\right).
\end{equation}

The ESDs are measured in the same $r_{\rm p}$ bins as the clustering measurement, namely 20 logarithmic bins ranging from $1\, h^{-1}\rm{Mpc}$ to $100\, h^{-1}\rm{Mpc}$. Since only large-scale ESD signals are considered in the subsequent analysis, it is deemed safe to use source galaxies with $z_s > z_l + 0.25$ to alleviate the dilution caused by unlensed foreground or member galaxies due to incorrect photo-$z$. However, the ESD signals might still suffer dilution due to the inaccurate lens and source galaxies redshift distribution. 


We propose a method to reduce the dilution utilizing cross-correlations between spectroscopic redshift galaxies and photometric redshift galaxies (similar to the concept of \cite{Newman2008ApJ}). This method is able to recover the joint probability distribution function of spectroscopic and photometric redshifts by using the photometric redshift probability distribution functions at each fine spectroscopic redshift bin. Through Bayes' theorem, we can calculate the dilution ratio that links the measured ESD to its theoretical value in the {\tt Fourier\_Quad} statistic approach. The details of the cross-correlation method will be presented in a forthcoming paper (Li et al. prep.). For our lens and source samples, the correction in ESD is on the order of a few percent.

After the correction, we further assess the impact of an inaccurate lens/source galaxy $n(z)$ on our ESD measurements by comparing the results obtained with a larger source-lens photo-$z$ cut ($z_s > z_l + 0.35$, Fig.~\ref{fig:lens_sys_dz_cut}).  ESD measurements in most redshift bins are consistent within 1$\sigma$ (2$\sigma$ in bin $0.5 < z_{\rm photo} < 0.7$). This test indicates that our results should be reliable (less so for the bin $0.5 < z_{\rm photo} < 0.7$).

Finally, we present the B-mode and ESD measurements from a random catalog in Appendix~\ref{subsec:appendix2}, as another systematic test for our shear measurement. 
We have subtracted the ESD of the random sample from that of the lens sample to correct the potential systematics.
The results indicate that no systematic bias is detected within the 2$\sigma$ level.

It is worth noting that the intrinsic alignment effect is likely not significant in our ESD measurements, due to the large source-lens separation cut in redshift, the large transverse separation, and the de-blending algorithm in the {\tt Fourier\_Quad} pipeline (e.g. \cite{Yao2020, Yao2023}). The roughly consistent ESDs seen in the test with a larger source-lens cut in redshift ($z_s > z_l + 0.35$) in Fig.~\ref{fig:lens_sys_dz_cut} indicate that the intrinsic alignment effect is probably safe to ignore in our measurements.

\begin{table*}[t]
\footnotesize
\caption{
Lens and source information and parameters constraints. 
Columns 1-8 list the redshift range of the lens sample,
the lower limit of absolute magnitude in the $z$-band (${M^{0.5}_{\rm z}-5\log h}$) for lens samples, 
the galaxy number in each lens sample, 
the catastrophic outlier rate, the mean lens redshift, the mean source redshift, the multiplicative bias
and the lens luminosity slope at $z_{\rm mag} = 21$.
The 9th column lists the reduced best-fit $\chi2$. 
Columns 10-13 list the four free parameters modeled for each lens sample: effective
photo-$z$ uncertainty $\sigma_{z}$, linear galaxy bias $b_g$,  
matter fluctuation $\sigma_8$, and matter density $\Omega_m$.
The values represent the median, and the upper and lower limits
corresponding to the 84 and 16 percentiles of the parameter
values from the MCMC chain.
The last column is the derived parameter $S_8 \equiv \sigma_8 \sqrt{\Omega_m/0.3}$.
}
\begin{center}
\begin{tabular}{cccccccccccccc}
\toprule
$\boldsymbol{z_l}$ & 
$\boldsymbol{M^{0.5}_{\rm z}}$ &
$\boldsymbol{N_{gal,l} \times10^{6}} $ & 
$\boldsymbol{\eta\times10^{-2}}$ &
$\boldsymbol{\bar{z_{l}}}$ & $\boldsymbol{\bar{z_{s}}}$ & $\boldsymbol{m\times10^{-2}}$ & $\boldsymbol{\alpha_{\rm mag}}$ & $\boldsymbol{\chi_r^2}$ & $\boldsymbol{\sigma_{z}\times10^{-2}}$ & 
$\boldsymbol{b_g}$ & $\boldsymbol{\sigma_{8}}$ & $\boldsymbol{\Omega_{m}}$ & $\boldsymbol{S_8}$
\\\hline
[0.1, 0.3] & -20.0 & 4.76 & 6.0 & 0.23 & 0.67 & 2$\pm$3 & 3.0 & 0.94 & $1.2^{+0.02}_{-0.02}$ & $1.22^{+0.02}_{-0.02}$ & $0.72^{+0.03}_{-0.03}$ & $0.39^{+0.02}_{-0.01}$ & $0.82^{+0.03}_{-0.03}$ \\\relax
[0.2, 0.4] & -20.0 & 6.82 & 5.2 & 0.32 & 0.74 & -4$\pm$4 & 2.8 & 0.91 & $1.6^{+0.02}_{-0.02}$ & $1.18^{+0.02}_{-0.02}$ & $0.80^{+0.02}_{-0.02}$ & $0.35^{+0.01}_{-0.01}$ & $0.87^{+0.02}_{-0.02}$ \\\relax
[0.3, 0.5] & -20.5 & 7.33 & 3.6 & 0.41 & 0.82 & -4$\pm$5 & 2.7 & 1.06 & $1.9^{+0.03}_{-0.03}$ & $1.20^{+0.02}_{-0.02}$ & $0.81^{+0.03}_{-0.03}$ & $0.34^{+0.01}_{-0.01}$ & $0.86^{+0.03}_{-0.03}$ \\\relax
[0.4, 0.6] & -21.0 & 5.46 & 3.4 & 0.50 & 0.88 & -5$\pm$6 & 2.6 & 1.22 & $1.6^{+0.02}_{-0.02}$ & $1.24^{+0.03}_{-0.02}$ & $0.84^{+0.03}_{-0.03}$ & $0.34^{+0.01}_{-0.01}$ & $0.89^{+0.04}_{-0.04}$ \\\relax
[0.5, 0.7] & -21.0 & 7.12 & 3.1 & 0.61 & 0.96 & -11$\pm$9 & 2.6 & 1.39 & $1.8^{+0.02}_{-0.02}$ & $1.27^{+0.03}_{-0.03}$ & $0.84^{+0.05}_{-0.05}$ & $0.34^{+0.01}_{-0.01}$ & $0.90^{+0.05}_{-0.06}$ \\\relax
[0.6, 0.8] & -21.5 & 3.97 & 2.8 & 0.70 & 1.02 & -33$\pm$15 & 2.5 & 1.92 & $2.0^{+0.03}_{-0.03}$ & $1.24^{+0.05}_{-0.04}$ & $0.97^{+0.07}_{-0.08}$ & $0.33^{+0.01}_{-0.01}$ & $1.02^{+0.08}_{-0.09}$ \\\relax
[0.7, 0.9] & -22.0 & 1.87 & 1.8 & 0.81 & 1.10 & -45$\pm$32 & 2.4 & 0.80 & $2.3^{+0.05}_{-0.05}$ & $1.38^{+0.16}_{-0.08}$ & $0.91^{+0.13}_{-0.20}$ & $0.35^{+0.01}_{-0.02}$ & $0.98^{+0.16}_{-0.22}$ \\
\bottomrule
\end{tabular}
\end{center}
\label{table:sample_info}
\end{table*}

\section{Joint modeling of galaxy-galaxy lensing and galaxy clustering}
\label{sec:method}

A combination of galaxy-galaxy lensing $\Delta\Sigma^{\rm obs}(\rp)$ and projected 2PCFs $w_{\rm p}^{\rm obs}(r_{\rm p})$ can effectively break the degeneracy between galaxy bias and cosmological parameters. 
To obtain unbiased cosmological parameters,
one may need to address two important questions: 1) At which scales would be considered free from complex nonlinear physics? 2) In perturbation theory-based methods, is the linear bias model sufficient, or should one introduce higher-order bias terms for unbiased parameter constraints? 
\cite{Sugiyama2020} investigated these questions using HSC-SDSS mock catalogs and found that as long as the scale cuts satisfy $\rp > 12~h^{-1}\rm Mpc$ and $8~h^{-1}\rm Mpc$ for $\Delta\Sigma^{\rm obs}(\rp)$ and $w_{\rm p}^{\rm obs}(r_{\rm p})$,
respectively, combining linear bias and nonlinear power spectrum can recover $\sigma_8$ and $\Omega_m$ within a 68\% credible interval. They also discovered that including higher-order bias contributions would slightly bias the parameter inference. Subsequently, they applied the minimal bias model to the HSC$\times$BOSS galaxy-galaxy weak lensing and BOSS galaxy clustering, obtaining reasonable cosmological constraints\cite{Sugiyama2022}, see also \cite{Shao2023}. Inspired by these studies, we adopt the minimal bias model and set the scale cuts at $r_{\rm p} = [8, 70]~h^{-1}{\rm Mpc}$ for $w_{\rm p}^{\rm obs}(r_{\rm p})$ and $[12, 70]~h^{-1}{\rm Mpc}$ for $\Delta\Sigma^{\rm obs}(\rp)$, respectively. The upper boundaries are chosen to avoid the effect from Baryon Acoustic Oscillations (BAO).
Simulations \cite{Pandey2020} and observational studies \cite{Elvin-Poole2018, Porredon2021} have also explored incorporating higher-order bias contributions by pushing to smaller scales (e.g., $\rp \sim 6-8~h^{-1}\rm Mpc$).

In this section, we present our methodology for constraining four free parameters in each lens sample: linear galaxy bias ($b_g$), effective photo-$z$ uncertainty ($\sigma_z$), matter density ($\Omega_m$), 
and matter fluctuation ($\sigma_8$), by jointly modeling galaxy-galaxy lensing ($\Delta\Sigma^{\rm obs}(\rp)$) and two sets of projected 2PCFs ($w_{\rm p}^{\rm obs}(r_{\rm p}|\pi_{\max} = 50~h^{-1}\rm Mpc)$ and $w_{\rm p}^{\rm obs}(r_{\rm p}|\pi_{\max} = 100~h^{-1}\rm Mpc)$). Unlike most 3$\times$2pt studies that use either spectroscopic lens samples for accurate 2PCFs or photometric lens samples with angular clustering, we intend to infer galaxy bias by jointly modeling two sets of projected 2PCFs from photometric samples with different integration lengths. By assuming a Gaussian distribution for the photo-$z$ uncertainty in each lens sample (Fig.~\ref{fig:photoz_err}, also see \cite{Zhou2021}), we can obtain the intrinsic clustering and effective photo-$z$ uncertainty spontaneously. This clustering-derived photo-$z$ uncertainty can be further compared with $\sigma_{\rm NMAD}$, estimated from a weighted spectroscopic subsample, serving as a sanity test for the Gaussian photo-$z$ PDF assumption. We describe the model ingredients for galaxy-galaxy lensing in Section~\ref{subsec:esd_model} and for projected 2PCFs in Section~\ref{subsec:wp_model}.

\subsection{Modeling galaxy-galaxy lensing: 
Excessive Surface Density $\Delta\Sigma^{\rm model}(\rp)$}
\label{subsec:esd_model}
The measured galaxy-galaxy lensing (GGL) signal 
$\Delta\Sigma^{\rm obs}(\rp)$ can be modeled as the sum of two terms: 
the standard excess surface density
$\Delta\Sigma(\rp)$ that probes the matter distribution around lens 
galaxies, and a systematic contamination introduced by the common 
foreground matter between us and the lens, termed as cosmic 
magnification bias $\Delta\Sigma^{\rm mag}(\rp)$, 
\begin{equation}
\label{eq:esd_mdoel}
\Delta\Sigma^{\rm model}(\rp) = \Delta\Sigma(\rp)+\Delta\Sigma^{\rm mag}(\rp)
\end{equation}

The standard ESD $\Delta\Sigma(\rp)$ is defined by the difference between the
average surface density within the projected distance $\rp$, $\Sigma(\leq\rp)$, and the projected surface density at $\rp$, $\Sigma(\rp)$, that is,
\begin{equation}
\label{eq:esd}
\Delta\Sigma(\rp) =\Sigma(\leq\rp)-\Sigma(\rp),
\end{equation}
The surface density $\Sigma(\rp)$ can be related to the galaxy-matter cross correlation function $\xi_{\rm gm}(r, z_l)$ at the lens redshift $z_l$ via the following equations:
\begin{equation}
\label{eq:sigatr}
\Sigma(\rp) = 2\, \overline{\rho}_{\rm m0} \int_{0}^{\infty} [1+\xi_{\rm gm}(\rp, \pi, z_l)] \,
{\rmd \pi }\,,
\end{equation}
and
\begin{equation}
\label{eq:siginr}
\Sigma(\leq \rp) = \frac{4\,\overline{\rho}_{\rm m0}}{\rp^2} \int_0^{\rp} y\,\,dy\,
 \int_{0}^{\infty} [1+\xi_{\rm gm}(y, \pi, z_l)]\, 
{\rmd \pi }\,,
\end{equation}
where $\overline{\rho}_{\rm m0}$ is the mean matter density of the Universe in the comoving coordinates, i.e, 
$\overline{\rho}_{\rm m0} = \Omega_m3H^2_0/(8\pi G)$ . 
The galaxy-matter cross-correlation function 
is defined with respect to the dark matter 
auto-correlation $\xi_{\rm mm}(r)$ by 
\begin{equation}
\xi_{\rm gm}(r, z_l)= b_g(r)r_{gm}(r)~\xi_{\rm mm}(r, z_l)\,.
\end{equation}
where $b_g(r)$ is the scale-dependent lens galaxy bias with respect to dark matter.  
And $r_{\rm gm}(r)$ is the cross-correlation coefficient
between the galaxies and dark matter, 
\begin{equation}
\label{eq:ccgm}
r_{\rm gm}(r) = \frac{\xi_{\rm gm}(r)}{\sqrt{\xi_{\rm mm}(r) \xi_{\rm gg} (r)} }\,.
\end{equation}
At the scales adopted for modeling, $r_{\rm gm}$ 
can be safely considered as unity, that is,
$r_{\rm gm} = 1$ \cite{Baldauf2010, Mandelbaum2013, Elvin-Poole2018, Singh2020, Sugiyama2020, Sugiyama2022}.
In the minimal bias model, $b_g(r)$ can be treated as an unknown scale independent constant, i.e.
$b_g(r) = b_g$, which is one of the free parameters in our model.

The matter between us and lens 
changes both source galaxy shapes and lens number densities, 
which contributes an extra correlation signal, termed as cosmic 
magnification bias $\Delta\Sigma^{\rm mag}(\rp)$
\footnote{
We ignore the magnification of source galaxy number densities 
by the matter around lens because it is a higher-order correlation 
compared to the magnification bias $\Delta\Sigma^{\rm mag}(\rp)$ 
considered in the text, also see \cite{Unruh2020, Miyatake2021}.}. 
It was used to be ignored in the GGL modeling, due to its tiny contribution. 
However, recent studies found that magnification bias
could account for $5\%\sim20\%$ signals from the measured GGL
(i.e. \cite{Unruh2020, vonWietersheim-Kramsta2021}), depending on the luminosity of the lens sample and the redshift of the lens and the sources samples. 
Therefore, it is crucial to include the contribution from cosmic magnification, because otherwise the cosmological constraint will be biased.
We model $\Delta\Sigma^{\rm mag}(\rp)$ through the following equation
(also see Eq.23 in \cite{Miyatake2021} and
Eq.14 in \cite{Unruh2020}),
\begin{align}
\label{eq:esd_mag}
\Delta\Sigma^{\rm mag}(\rp) &\approx  2(\alpha_{\rm mag}-1)\frac{3H_0\Omega_m}{2c}
\int_0^{\bar{z_l}}dz\frac{H_0(1+z)^2}{H(z)(1+\bar{z_l})}\nonumber\\
& \times \frac{\chi^2(\bar{\chi_l}-\chi)(\bar{\chi_s}-\chi)}{\bar{\chi_l}^2(\bar{\chi_s}-\bar{\chi_l})}\nonumber\\
& \times \overline{\rho}_{m0}
\int\!\frac{k\mathrm{d}k}{2\pi}P_{\rm mm}^{\rm NL}\!(k;z)J_2\left(k\frac{\chi}{\bar{\chi_l}}\rp\right),
\end{align}
where $c$ is the vacuum speed of light and $J_2(x)$ is the second-order Bessel function of the first kind. The estimated luminosity slopes $\alpha_{\rm mag}$ for each lens sample are listed in Table \ref{table:sample_info}. The determination of the exact values of $\alpha_{\rm mag}$ is not trivial with the complex sample selection functions adopted in this work (e.g. \cite{vonWietersheim-Kramsta2021}).
\cite{Miyatake2021} found that the inferred values of $\sigma_8$ and $\Omega_m$
remain almost the same regardless of treating them as free parameters or adopting fixed estimated values. 
For simplification, we use $\alpha_{\rm mag}$  estimated from luminosity function for each lens sample in our modeling.
We approximate the lens and source galaxies redshift
as their mean redshift $\bar{z_{l}}$ and $\bar{z_{s}}$ and their corresponding comoving distance $\bar{\chi_l}$ and $\bar{\chi_s}$.
The theoretical templates $\xi^{\rm NL}_{\rm mm}(r, z_l)$ and $P_{\rm mm}^{\rm NL}\!(k)$ are \texttt{halofit} \cite{Takahashi2012} nonlinear correlation and power spectrum for a given cosmology returned from python package \texttt{pyccl} \cite{Chisari2019}. 
The last
line in Eq.\ref{eq:esd_mag} (the inverse Hankel transform) is performed with Python package \texttt{ Hankel} \cite{Murray2019}.

\subsection{Modeling galaxy clustering: two sets of projected 2PCFs 
$w_{\rm p}^{\rm model}(r_{\rm p}|\pi_{\max} = 50~h^{-1}\rm Mpc)$ and 
$w_{\rm p}^{\rm model}(r_{\rm p}|\pi_{\max} = 100~h^{-1}\rm Mpc)$ }
\label{subsec:wp_model}
To model the projected 2PCFs from photometric galaxy samples with photo-$z$,
we start from the definition,
\begin{equation}
\label{eq:wrp}
w^{\rm model}_p(r_{\rm p}|\pi_{\max}) = 
2 \int_{0}^{\pi_{\max}} \xi^{{\rm photo-}z}_{\rm gg}(r_{\rm p}, \pi) \,{\rmd \pi }\,,
\end{equation}
where the superscripts ``photo-$z$'' 
in $\xi^{{\rm photo-}z}_{\rm gg}(r_{\rm p}, \pi)$ denotes that 
pair-counting is done in the photo-$z$ space, which is further 
related to the 2PCFs in redshift space 
$\xi^{\rm s}_{\rm gg}(r_{\rm p}, \pi)$ through
\begin{equation}
\xi^{{\rm photo-}z}_{\rm gg}(r_{\rm p},\pi) = (1-\eta)^2\int^{\infty}_{-\infty}\xi^{\rm s}_{\rm gg}(r_{\rm p},\pi-R)P(R){\rm d}R\,,
\label{eq:streaming}
\end{equation}
where the impact of the photo-$z$ uncertainties on the 2PCFs is modeled by two
components: photo-$z$ outliers with random photo-$z$ distribution and non-outliers with a Gaussian photo-$z$ distribution. It is important to note
that the meaning of outliers in this context is distinct from the one in Section~\ref{subsec:sample_photoz}, which is now defined from the perspective of a Gaussian distribution, i.e. $|z_{\rm photo}-z_{\rm spec}|/(1+z_{\rm spec}) > 3\sigma_{\rm NMAD}$.
For nonoutliers, the PDF of the difference between the
photo-$z$ derived pair separation and the spectroscopic redshift derived pair separation is as follows.
\begin{align} 
  P(R) &= \frac{1}{\sqrt{2\pi}\sigma_{\rm R}}\exp\left(-\frac{R^2}{2\sigma_{\rm R}^2}\right), \\ \nonumber
  \sigma_R &= \frac{\sqrt{2}c\sigma_z(1 + z_{\rm spec})}{H_0 E(z_{\rm spec})}.
\label{eq:Gaussian_photoz}
\end{align}
where $E(z)=\sqrt{\Omega_{m}(1+z)^{3}+\Omega_{\Lambda}}$, and 
$\sigma_z$ is the effective photo-$z$ uncertainty of lens sample.
The scale factor $(1-\eta)^2$ accounts for the missing galaxy pairs
due to photo-$z$ outliers, assuming that
these catastrophic outliers are randomly
distributed in the redshift direction (see the derivation in Eq. 9 in \cite{Wang2019}).
The percentage of catastrophic outliers $\eta$ is directly estimated from Fig.~\ref{fig:photoz_err}.


Note that in Eq.\ref{eq:wrp}, the summation is obtained
by integrating to a finite light-of-sight separation $\pi_{\max}$, 
implicitly ignoring the peculiar velocities on scale $r>\pi_{\max}$.
For $\pi_{\max} = 100~h^{-1}\rm Mpc$, 
as demonstrated in \cite{vandenBosch2013}, the so-called residual redshift-space
distortion (RRSD) could cause a 15\% correction at $r_{\rm p} \sim 30~h^{-1}\rm Mpc$.
To take this effect into account,
we take the approach as laid out by Eq.51-56 in \cite{vandenBosch2013}, which assumes that peculiar velocities in large scale (i.e. $r> \pi_{\max}$) can be
modeled by linear perturbation theory. Therefore, the galaxy 2PCFs in redshift space
can be written as (e.g. \cite{Kaiser1987})
\begin{equation}
    \xi^{\rm s}_{\rm gg}(r_{\rm p},\pi) = \xi_0 (s)\mathcal{P}_0(\mu)+\xi_2(s)\mathcal{P}_2(\mu)+\xi_4 \mathcal{P}_4(\mu) ,
\end{equation}
where $s = \sqrt{r^2_{\rm p}+ \pi^2}$ is the galaxy separation in redshift-space and $\mu$ is
the cosine of the angle between the separation $s$ and line-of-sight direction, i.e. 
$\mu = \pi/s$.
$\mathcal{P}_{0,2,4} (x) $ is the $0^{\rm th}$, $2^{\rm nd}$, $4^{\rm th}$ Legendre polynomials, respectively.
The multipoles $\xi_{0,2,4}$ are related to the real space 2PCFs $\xi_{\rm gg}(r)$ by
\begin{equation}
\label{eq:xi_multi}
    \xi_0(r) = \left( 1 + \frac{2}{3}\beta+\frac{1}{5}\beta^2\right)\xi^{\rm lin}_{\rm gg}(r),
\end{equation}
\begin{equation}
    \xi_2(r) = \left( \frac{4}{3}\beta+\frac{4}{7}\beta^2\right)[\xi^{\rm lin}_{\rm gg}(r)-3J_3(r)],
\end{equation}
\begin{equation}
    \xi_4(r) = \frac{8}{35}\beta^2\left[\xi^{\rm lin}_{\rm gg}(r) + \frac{15}{2}J_3(r)- \frac{35}{2}J_5(r)\right],
\end{equation}
where
\begin{equation}
\label{eq:xi_gg}
\xi^{\rm lin}_{\rm gg}(r) = b_g\xi^{\rm lin}_{\rm mm}(r),
\end{equation}
\begin{equation}
\label{eq:J_n}
    J_n(r) = \frac{1}{r^n}\int^r_0\xi^{\rm lin}_{\rm gg}(y)y^{n-1}dy,
\end{equation}
and 
\begin{equation}
    \beta = \frac{1}{b_g}\left( \frac{d \rm lnD}{d \rm lna}\right)_z.
\end{equation}
where $a$ is the scale factor defined as $a\equiv1/(1+z)$ and $D$ is the linear growth rate. 
To be consistent, 
we replace the linear matter 2PCFs $\xi^{\rm lin}_{\rm mm}(r)$ in Eq.\ref{eq:xi_multi}-Eq.\ref{eq:J_n}
with nonlinear 2PCFs $\xi^{\rm NL}_{\rm mm}(r)$ as in the previous subsection.
This replacement could remove possible over-correction at small scales.

\begin{figure*}
\centering
\includegraphics[width=\textwidth]{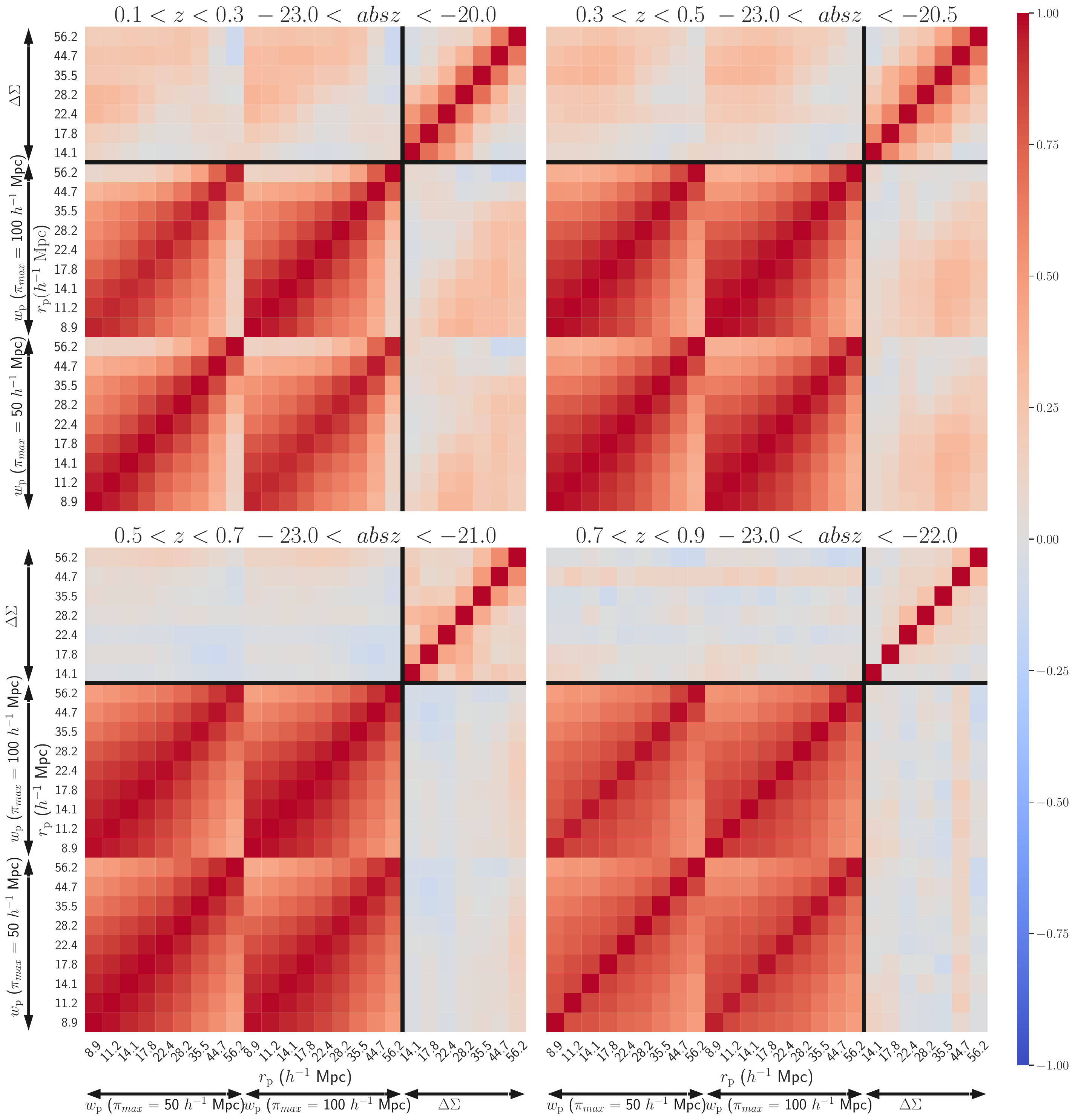}
\caption{
Normalized covariance matrices of observable vector $\mathbf X^{\rm obs}_{p} \equiv [w^{\rm obs}_{\rm p}(\pi_{\rm max}=50~h^{-1}{\rm
Mpc}), ~ w^{\rm obs}_{\rm p}(\pi_{\rm max}=100~h^{-1}{\rm Mpc}), ~ \Delta\Sigma^{\rm obs}]$ for 4 out of 7 lens samples that range from low to high redshift. 
The redshift and luminosity information, in the same format as those in Fig.~\ref{fig:photoz_err}, is indicated on top of the quartet.
{\it Clustering}: Within $w^{\rm obs}_{\rm p}(\pi_{\rm max}=50~h^{-1}{\rm Mpc})$
(or $w^{\rm obs}_{\rm p}(\pi_{\rm max}=100~h^{-1}{\rm Mpc})$),
the strong correlations between radial bins are
expected, reflecting the impact of large-scale modes.
Since $w^{\rm obs}_{\rm p}(\pi_{\rm max}=100~h^{-1}{\rm Mpc})$ contains
the information of $w^{\rm obs}_{\rm p}(\pi_{\rm max}=50~h^{-1}{\rm Mpc})$, 
strong correlations are expected between them.
{\it Galaxy-galaxy lensing}:
in the lowest redshift lens sample, there exist considerable
correlations between neighboring bins. 
{\it Clustering $\times$ Galaxy-galaxy lensing}:
at low redshift, small-scale clustering is correlated with galaxy-galaxy 
lensing at all scales, while large-scale clustering is much less
correlated with GGL. 
The correlation strength also depends on the redshift. 
For the lens sample with the highest redshift,
the clustering is almost uncorrelated with GGL at all scales.
}
\label{fig:cov}
\end{figure*}



\begin{figure*}[!t]
\centering
\includegraphics[width=0.96\textwidth]{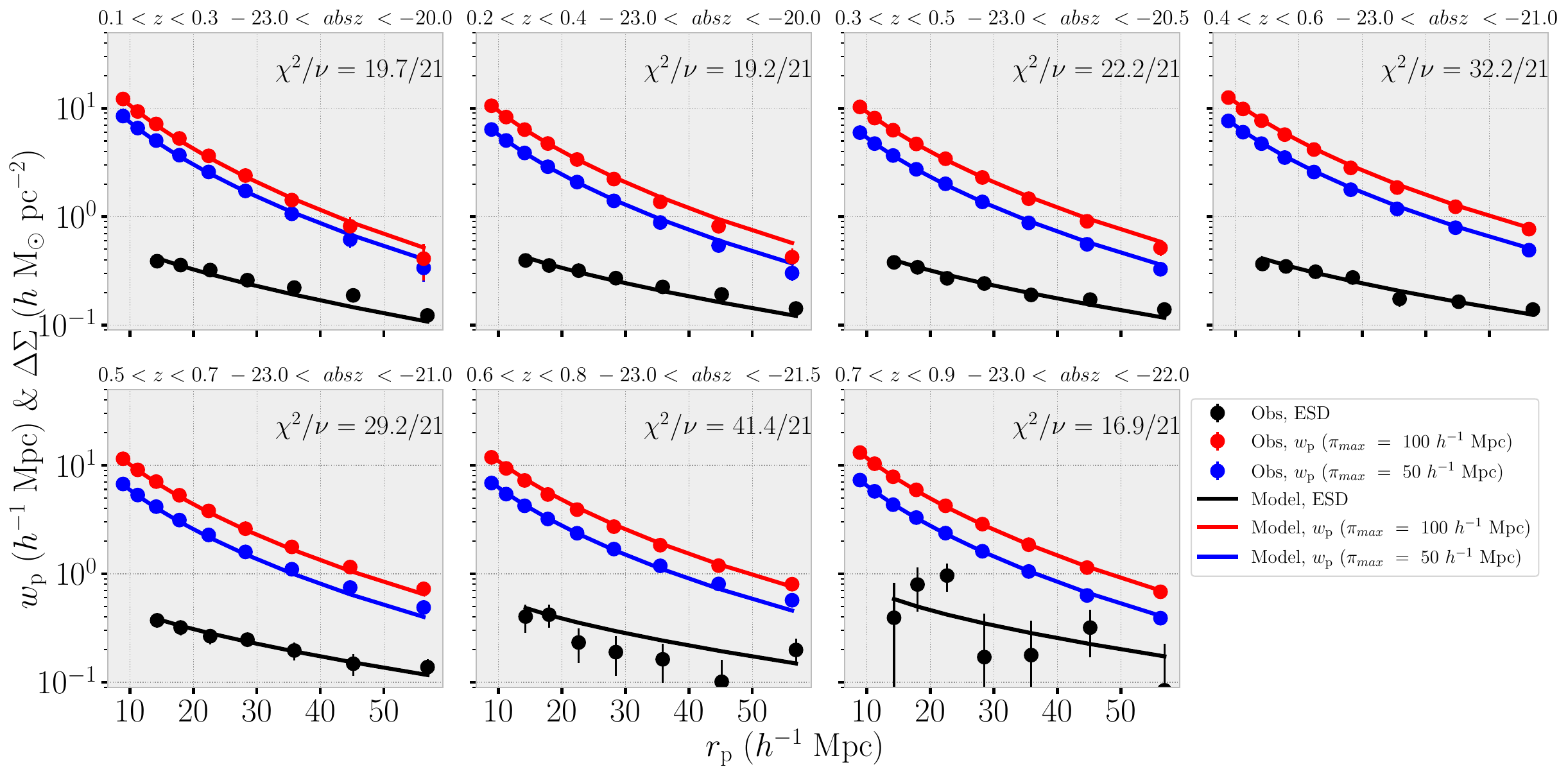}
\caption{
Observation (symbols) VS. best-fit model (lines) for seven lens samples.
The blue and red circles represent the imaging-systematics-mitigated 
projected 2PCFs $w_{\rm p}^{\rm obs}(r_{\rm p}|\pi_{\max} = 50~h^{-1}\rm Mpc)$ and 
$w_{\rm p}^{\rm obs}(r_{\rm p}|\pi_{\max} = 100~h^{-1}\rm Mpc)$, respectively.
The blue and red lines are their corresponding best-fit model.
The ESD measurements (best-fit model) 
are shown as black circles (lines).
The degree of freedom $\nu = 21$ is simply calculated from the number of data points 
$N_{\rm data} = 25$ minus the number of free parameters $N_{\rm para} = 4$.
Note that due to uncertain redshift distribution and limited source galaxies, the lensing measurements in the highest three redshift bins (bottom row) are of less trustworthy. 
}
\label{fig:obsVSmodel}
\end{figure*}

\subsection{Model Fitting}

We set out our model to constrain the four free parameters for each lens sample: $\Omega_m$, $\sigma_8$, $b_g$, and $\sigma_z$, by maximizing the posterior distribution
\begin{align}
    P_{\rm posterior}(\Omega_m, \sigma_8, b_g, \sigma_z) = P_{\rm prior}(\Omega_m, \sigma_8, b_g,\sigma_z )\nonumber\\
    \times \mathcal L(\mathcal D|\Omega_m, \sigma_8, b_g, \sigma_z),
\end{align}
where the likelihood is
\begin{align}
\label{eq:likelihood}
\log\mathcal L & \propto ({\mathbf X_{p}^{\rm obs}}-{\mathbf X_{p}^{\rm model}})^{\rm T}{\mathbf C^{-1}}{({\mathbf X_{p}^{\rm obs}}-{\mathbf X_{p}^{\rm model}})}, 
\end{align}
where $\mathbf X^{\rm obs}_{p}$ ($\mathbf X^{\rm model}_{p}$) represents the 
concatenated observable (model) vector, which consists of 
two sets of projected 2PCFs and one set of ESD measurements,
i.e., $\mathbf X^{\rm obs (model)}_{p} =[w^{\rm obs (model)}_{p}(\pi_{\rm max}=50~h^{-1}{\rm
Mpc}), ~ w^{\rm obs (model)}_{p}(\pi_{\rm max}=100~h^{-1}{\rm Mpc}), ~ \Delta\Sigma^{\rm obs (model)}]$.
Within the scale cuts for clustering $r_{\rm p} = [8, 70]~h^{-1}{\rm Mpc}$,
we have 9 data points for each $w_{\rm p}^{\rm obs}(r_{\rm p})$.
For galaxy-galaxy lensing $r_{\rm p} = [12, 70]~h^{-1}{\rm Mpc}$, 
we have 7 data points. 
In total, each lens sample provides $N_{\rm data} = 25$ data points. 
The corresponding 25$\times$25 covariance matrix, ${\mathbf C}$, is obtained through Jackknife re-sampling method with $N_{\rm jkf} = 200$. 
When inverting the covariance matrix, 
we apply the Hartlap correction
\cite{Hartlap2007} $(N_{\rm jkf} - N_{\rm data} - 2)/(N_{\rm jkf} - 1)$ to account for the bias. 
Fig.~\ref{fig:cov} showcases the covariance matrices for 4 out of 7 lens samples.
To sample the posterior distribution, we run affine-invariant ensemble samplers for Markov Chain Monte Carlo (MCMC, \cite{Goodman2010}) implemented in the Python package \texttt{ emcee} \cite{Foreman-Mackey2013}.
To minimize the influence of the prior, we set an uninformative
uniform prior distribution for all parameters:
$\Omega_m \sim [0.2, 0.6]$, $\sigma_8 \sim [0.5, 1.1]$, $b_g \sim [0.5, 4.5]$, and $\log\sigma_z \sim [-3, -1]$. For each lens sample, we run MCMC with 144 walkers and
30000 steps per walker for a total of 4320000 model evaluations.

We use the integrated autocorrelation time ($\tau$) provided within the {\tt emcee} to assess the convergence of the chains. 
The basic idea of integrated autocorrelation time is
to determine the steps required for the chains to be considered independent. 
The higher the number of effective independent samples ($N_{\rm chain}/\tau$), 
the fewer sampling errors present in the integrals computed using
the MCMC results. 
A reliable estimate is typically achieved when $N_{\rm chain}/\tau > 50$, 
which serves as the default convergence criterion in the {\tt emcee}.
We require all MCMC chains to reach this convergence threshold.
In the end, we have verified that each parameter converges to a fixed value after a certain number of iterations. 
For more in-depth information on autocorrelation analysis and convergence, we refer interested readers to this link \footnote{\url{https://emcee.readthedocs.io/en/stable/tutorials/autocorr/}} and the reference therein.

During the measurement of projected 2PCFs and galaxy-galaxy lensing signals, we assume Planck 2018 cosmology \cite{Planck2018} to calculate distances between galaxies. However, at high redshift (for example $z > 0.5$), the distance would change significantly if a different value of $\Omega_m$ were used. This introduces a systematic bias when comparing the measurements and models with varying $\Omega_m$. To address this issue, we employ the mapping method provided by \cite{More2013a} to eliminate the cosmological dependence of the measurements. In other words, for each MCMC model evaluation, we scale the model predictions to the Planck 2018 cosmology under which measurements are performed.

\begin{figure*}[!t]
\centering
\includegraphics[width=0.96\textwidth]{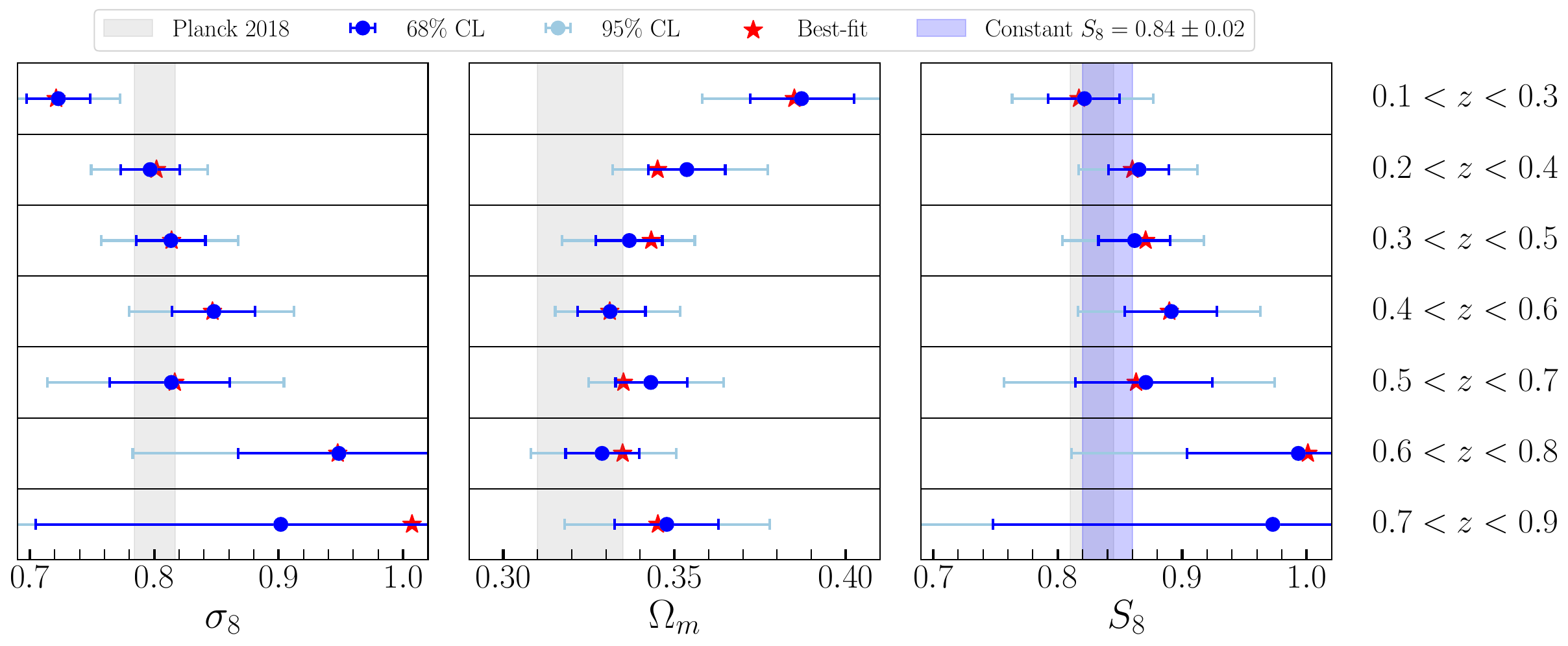}
\caption{Cosmological parameters constraints from seven lens samples.
The sample redshift is shown on the right.
For each column, the filled circles are the median values obtained from the marginalized posterior, and the thin and thick error bars denote the 68\% and 
95\% credible intervals, respectively. 
In addition to the median, we also plot the best-fit parameters. 
For comparisons, the vertical gray bands denote 68\% credible intervals from Planck 2018 \cite{Planck2018}.
The light blue band denotes $S_8 = 0.84\pm0.02$, fitting with a constant $S_8$ in the first and third redshift bins. We note that the number changes to $S_8 = 0.87\pm0.01$ if the second and fourth bins were used.
}
\label{fig:cosmos_parameter}
\end{figure*}

\section{Model Fitting Results}
\label{sec:results}
With the above parametrization, we model the clustering and lensing 
measurements for each lens sample. 
The MCMC chains of each sample have been checked to converge and
the posterior distributions of the model parameters are summarized in
Table~\ref{table:sample_info}. 

In this section, we present our modeling results. We begin by assessing the performance of the model through a comparison of the measurements and model predictions in section~\ref{subsec:obsvsmodel}. 
We then focus on the parameter constraints and the derived parameters $S_8$
in Section~\ref{subsec:para_constraint}. Finally, in Section~\ref{subsec:model_test}, we perform two robustness tests to validate our model assumptions.

\subsection{Observation versus model}
\label{subsec:obsvsmodel}

Fig.~\ref{fig:obsVSmodel} presents comparisons between measurements and model predictions for projected 2PCFs and ESD. In general, our best-fit models demonstrate reasonable agreement with the measurements. 
The models perform better in fitting clustering signals compared to lensing signals. Specifically, our models effectively reproduce the clustering at small scales and exhibit somewhat less accurate at larger scales. This could be attributed to potential over- or under-correction of imaging systematics mitigation. Further discussion on imaging systematics mitigation can be found in section~\ref{subsec:diss_clustering}.

Due to limited source galaxies ($z_s > z_l + 0.25$), the lensing measurements in the two highest redshift lens samples are more uncertain than those in the lower redshift bins. 
While for the redshift bin $0.5 < z < 0.7$, the lensing measurements seem to be sensitive to the source galaxies redshift distribution, as shown in Fig.~\ref{fig:lens_sys_dz_cut}.
In summary, the lensing measurements and the resulting cosmological parameters in the three highest redshift bins are less reliable, but we still keep the results for reference.

\subsection{Cosmological parameter constraints}
\label{subsec:para_constraint}

After marginalizing out the photo-$z$ uncertainties and galaxy bias, we present the constraints of the cosmological parameters $\sigma_8$ and $\Omega_m$ in Fig.~\ref{fig:cosmos_parameter}.
With increasing redshift, $\sigma_8$ increases, while $\Omega_m$ decreases. 
This pattern probably stems from the well-known degeneracy between these two parameters when combining clustering and galaxy-galaxy lensing signals.
Using the tomographic cross-correlation of DESI LRG and Planck CMB lensing, \cite{White2022} also observed an increase in $\sigma_8$ with lens redshift.

The combination of clustering and galaxy-galaxy lensing demonstrates
greater sensitivity to the derived structure growth parameter $S_8 \equiv \sigma_8 \sqrt{\Omega_m/0.3}$, shown in Fig.~\ref{fig:cosmos_parameter}. 
The constraining power of $S_8$ appears to decrease with increasing redshift, likely due to growing uncertainties in the lensing measurements. 
The best-fit values of $S_8$ suggest a mild increase trend with lens redshift, although with low statistical significance. The rise $S_8$ implies either improvements needed in measurements and modeling or new physics beyond the standard $\Lambda$CDM model. Investigating the exact causes requires careful control in our measurements and modeling. The numerical values of the constraints are listed in Table~\ref{table:sample_info}.
Note that our constraints on the galaxy linear biases ($\sim 1.3$) in the high redshift bins ($z>0.5$) are somewhat lower than LRG-like galaxies at similar redshift ($\sim 2$, see \cite{Zhou2021}). The underestimation is probably related to the rising $\sigma_8$ or $S_8$ associated with the ESD measurements in the high redshift bins where the uncertainties are relatively larger (see Appendix~\ref{subsec:appendix2}). 

\begin{figure}[H]
\centering
\includegraphics[width=0.45\textwidth]{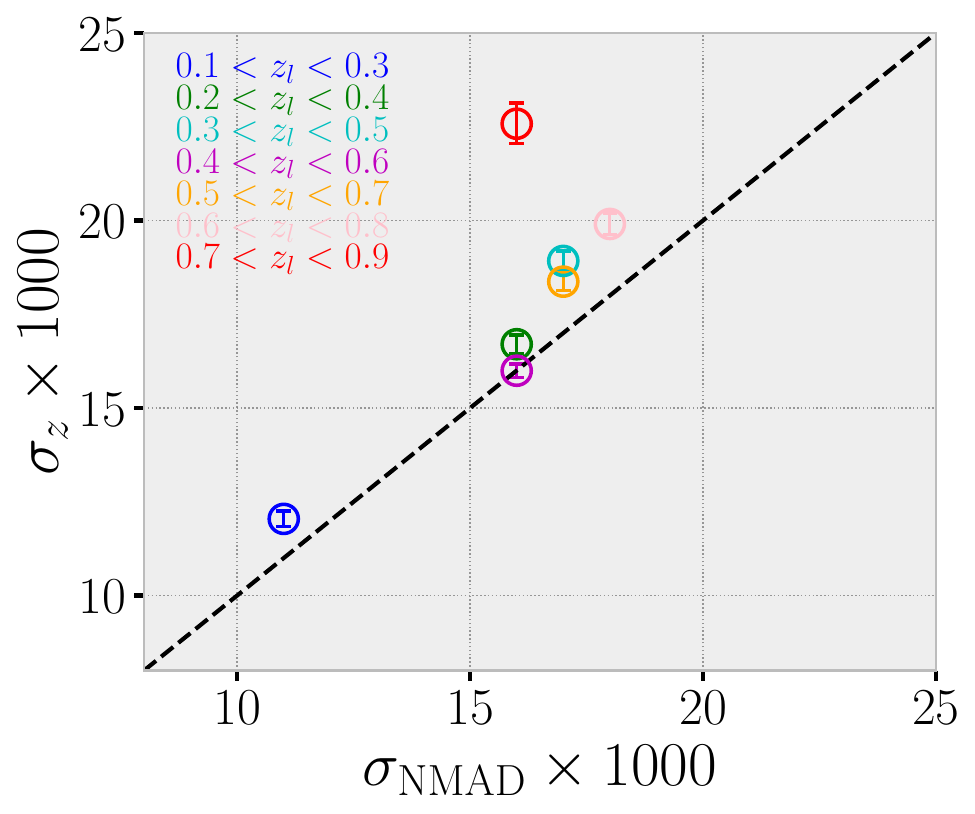}
\caption{
Comparison of effective photo-$z$ errors inferred from
clustering ($\sigma_z$, $y$-axis) and color-redshift relation
($\sigma_{\rm NMAD}$, $x$-axis) for seven lens samples.
The $\sigma_z$ are marginalized posterior median values
and the error bars denote the 68\% credible intervals (listed in Table ~\ref{table:sample_info}).
The normalized median absolute deviation $\sigma_{\rm NMAD}$
is estimated from the black histograms in Fig.~\ref{fig:photoz_err}.
The symbol colors represent the redshift bin listed in the upper left corner.
The black dashed line is the ``1:1'' line. 
The $\sigma_z$ is inferred from the large-scale structure, while
$\sigma_{\rm NMAD}$ is estimated from the galaxy color-magnitude relation.
These estimators utilize different information and therefore contain different systematics. 
However, the good agreement
suggests that our photo-$z$ Gaussian assumption is probably not
too off.
}
\label{fig:sigmaz_compare}
\end{figure}

\begin{figure*}[!t]
\centering
\includegraphics[width=0.96\textwidth]{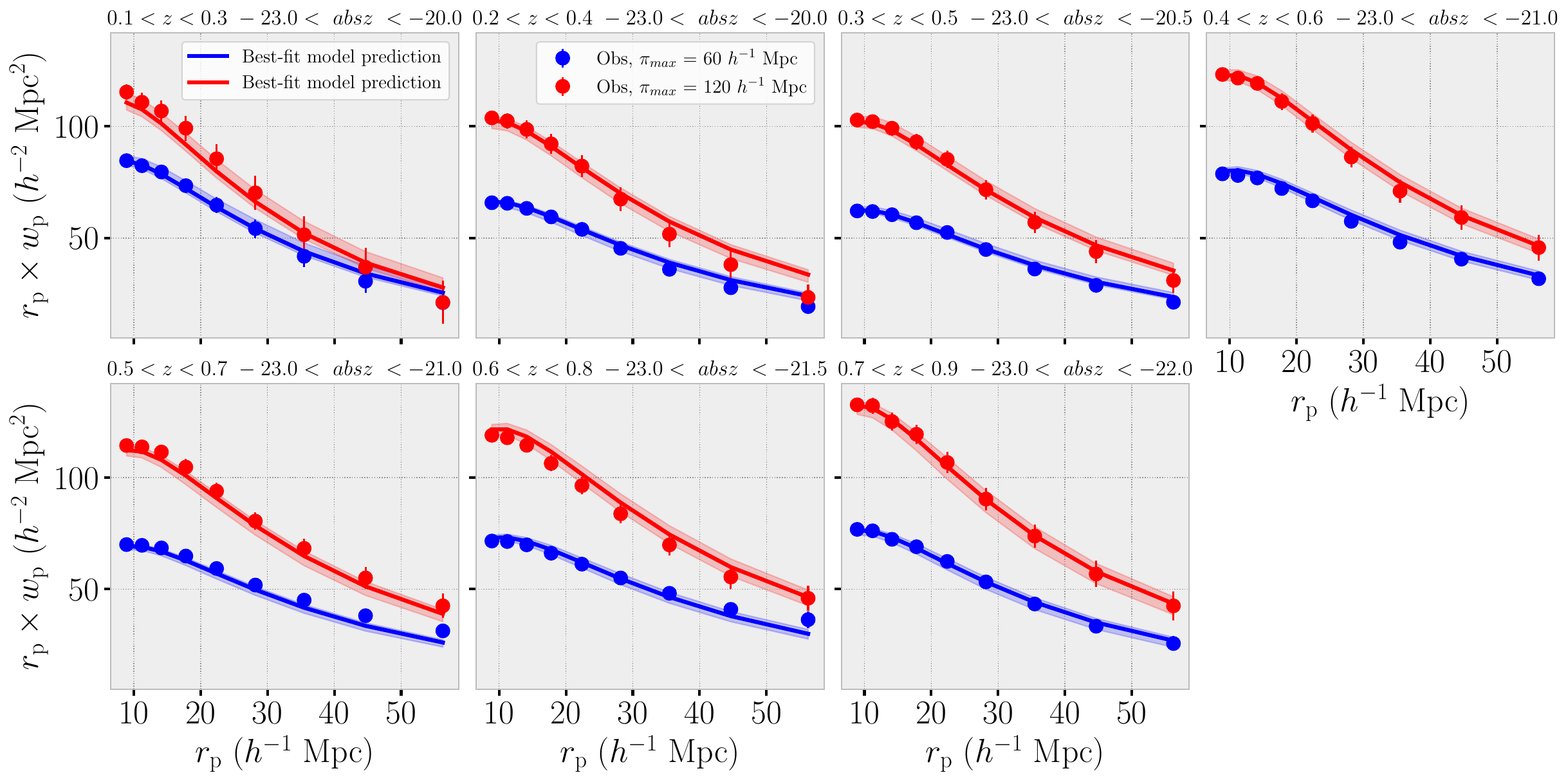}
\caption{
Model prediction VS. measured projected 2PCFs at different projection depths.
The points are measurements with error bars from jackknife. The solid curves are the predictions from best fit parameters. 
The bands are the [16th, 84th] percentile range of the predicted clustering signals.
We scale the measured 2PCFs by a factor of $1/(1-\eta)^2$ to compensate for the missing clustering strength due to photo-$z$ outliers.}
\label{fig:model_prediction}
\end{figure*}

When fitting with a constant $S_8$, we obtained $S_8 = 0.84\pm0.02$ ($S_8 = 0.87\pm0.01$), using the first (second) and third (fourth) redshift bin.
The combined constraint is consistent with the Planck 2018 results within 1$\sigma$, but significantly higher than latest weak lensing 3$\times$2pt probes by 2-5$\sigma$, i.e. DES-Y3 \cite{DES_Y3} ($S_8 = 0.776\pm0.017$) , KiDS-1000 \cite{Heymans2021} ($S_8 = 0.766^{+0.02}_{-0.014}$), and HSC-Y3 \cite{HSC_Y3} ($S_8 = 0.775^{+0.043}_{-0.038}$).
As we mentioned, more systematic tests on our measurements and modeling are needed before accessing the tensions
between our constraints and results from weak lensing surveys.

\subsection{Model tests}
\label{subsec:model_test}

In order to model the projected 2PCFs measured with photo-$z$, we make the assumption that the difference between the lens galaxy's photometric redshift and
its true redshift follows a Gaussian distribution. 
The width of this distribution represents the effective photo-$z$ uncertainty for the lens samples. By comparing the differences between two sets of 2PCFs that integrate to two different light-of-sight depths $\pi_{\max}$, we can directly assess this uncertainty. 
Although Fig.~\ref{fig:photoz_err} supports our assumption,
it is still very restrictive to assume the Gaussian photo-$z$ PDF. 
As a sanity test, we compare the effective photo-$z$ error derived from clustering ($\sigma_z$) with that derived from the galaxy magnitude-color information ($\sigma_{\rm NMAD}$) . 
The $\sigma_z$ is inferred from the large-scale structure, while
$\sigma_{\rm NMAD}$ is estimated based on the galaxy color-magnitude-redshift relation. These estimations utilize different information and each have their own model assumptions, which may suffer different systematic errors.

The comparison between two estimations of the photo-$z$ uncertainty is presented in
Fig.~\ref{fig:sigmaz_compare}.
The small differences (on the order of $\sim 0.001$) 
suggest that our assumption of a Gaussian photo-$z$ distribution is likely not far off.

With our chosen model parameters, we can predict the projected 2PCFs for any integration depth.
We compare the prediction with measurements for integration depths that were not used in the fitting. The comparison is presented in Fig.~\ref{fig:model_prediction}. In general, our model predictions align reasonably well with the measurements.

\section{Discussions}
\label{sec:diss}

In this section, we discuss possible caveats
in our measurements and modeling that may impact our final parameter constraints.

\subsection{Possible caveats in clustering measurements and modeling}
\label{subsec:diss_clustering}

The possible caveats in the clustering part lie in the following procedures: imaging systematic correction, compensation for photo-$z$ outliers, and Gaussian photo-$z$ PDF assumption.

\begin{figure}[H]
\centering
\includegraphics[width=0.48\textwidth]{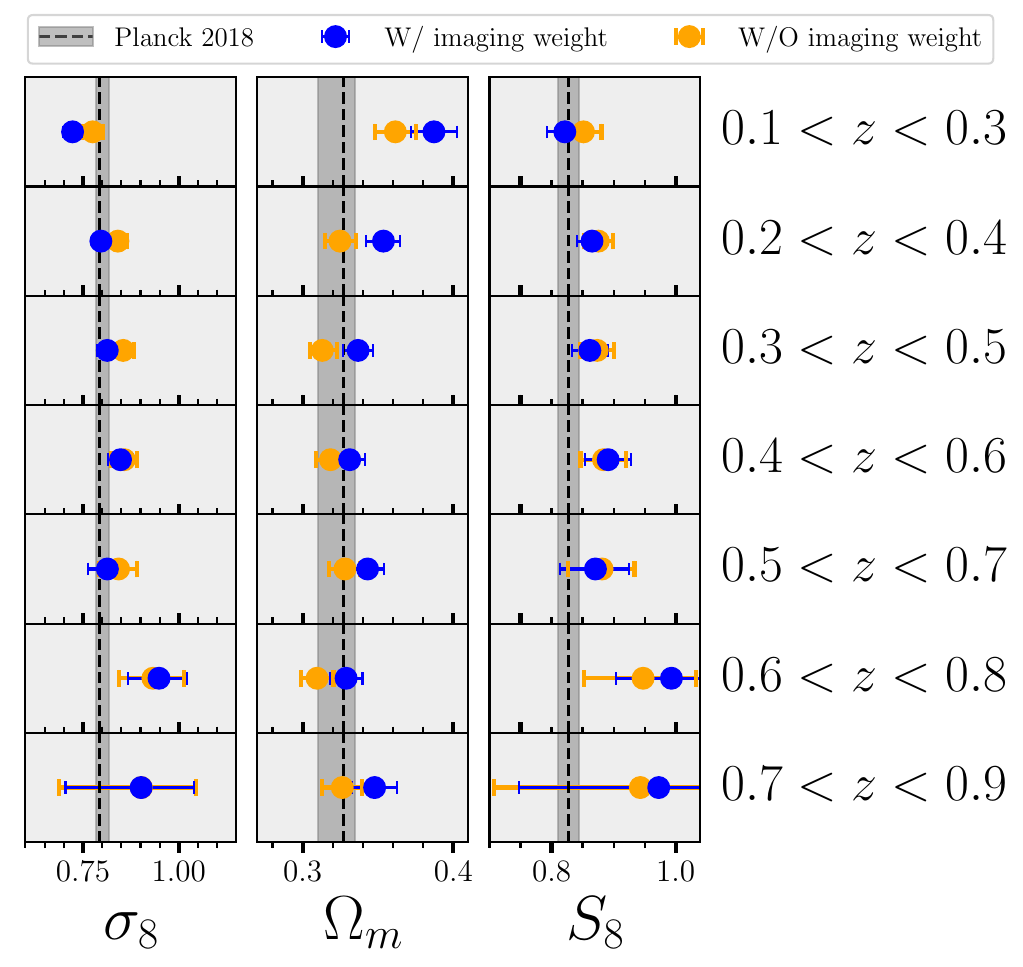}
\caption{
Comparison of cosmological parameters constraints with (blue, fiducial) 
and without (orange) imaging systematics correction in the lens sample.
}
\label{fig:systematics_compare}
\end{figure}

In Section~\ref{subsec:imaging}, we employ the well-developed 
code {\tt regressis} to mitigate spurious fluctuations in the lens samples. 
After correction, the lens samples no longer exhibit any dependence on the imaging quantities (Fig.~\ref{fig:imaging_systematics}). We have also verified that the corrected samples do not show correlation with the imaging maps.
However, recent research (e.g. \cite{Rodriguez-Monroy2022}) suggests that the machine learning-based mitigation method may excessively suppress clustering signals. To what extent the systematics correction would propagate to cosmological parameter constraints, we compare the inference with and without imaging systematics correction in clustering, as shown in Fig.~\ref{fig:systematics_compare}.

As illustrated in Fig.~\ref{fig:wp_w_sys_weight}, imaging correction has more significant effects on the clustering amplitude at large scales compared to small scales. 
This explains why $\Omega_m$ changes significantly with or without corrections since
$\Omega_m$ is more sensitive to the overall shape of the matter power spectrum. 
However, $S_8$ exhibits relatively small changes, indicating that
$S_8$ is probably more robust against systematics. 
Therefore, obtaining accurate mitigation of imaging systematics is crucial to determine the exact values of $\sigma_8$ and $\Omega_m$. In principle, it can be tested by applying {\tt regressis} to realistic mock lens samples, which will be carried out in future work.

When modeling the projected 2PCFs (see section \ref{subsec:wp_model}), 
we assume that the photo-$z$ of most lens galaxies follow a Gaussian distribution and that the rest a few percent of galaxies (i.e. catastrophic outliers) have a randomly distributed photo-$z$.  
The results shown in Figs.~\ref{fig:photoz_err}, ~\ref{fig:sigmaz_compare} and ~\ref{fig:model_prediction} suggest that both assumptions are likely not too bad for our lens galaxies, since they are LRG-like galaxies with relatively reliable photo-$z$ measurements, though the derived linear bias is somewhat lower than expected.
However, a more sophisticated analysis should incorporate photo-$z$ bias, skewness, or
kurtosis as nuisance parameters to describe the shape of the photo-$z$ distribution, 
which we plan to include in future studies. 

To compensate for the clustering strength due to the random distributed catastrophic outliers, we scale the model prediction by a factor of $(1-\eta)^2$ in Eq.~\ref{eq:streaming}.
The percentage of catastrophic outliers, $\eta$, is estimated by re-weighting the
spectroscopic subsample to match the distribution of the photometric sample in the color-magnitude space. We use the $k$-nearest neighbors method ($k$NN, \cite{Lima2008}) to derive the weights, where $k=1$ is adopted in our case. 
However, determining the exact number of neighbors for a fair estimation of $\eta$ is not trivial, and therefore the percentage of outliers may suffer some uncertainty.  
It should be noted that the weights can also be derived using other methods, for example, the self-organizing map (SOM) method (e.g. \cite{Hikage2019}). 
During the fitting process, we keep the outlier percentage $\eta$ fixed, rather than treating it as a nuisance parameter. The reason is that $\eta$ is in complete degeneracy with linear bias and $\sigma_8$. Leaving it free would significantly degrade the constraints. 
However, in our approach, an incorrect estimate of $\eta$ may lead to a biased $\sigma_8$. 
The impact of outliers on cosmological constraints deserves further investigation.

\subsection{Possible caveats in lensing measurements and modeling}
\label{subsec:diss_lensing}


We used a spec-$z$ sample to estimate the correction for photo-$z$ dilution in ESD measurements. This sample is divided into fine spec-$z$ bins ($\Delta z=0.01$) (Li et al. in prep.). The accuracy of recovery is limited by the size of the available spec-$z$ sample. For instance, at $z \sim 0.5$, more than $20000$ spec-$z$ galaxies are present in the footprint of our shear catalog, while for the higher spec-$z$ bins ($z>1$), only a few thousand spec-$z$ galaxies are available. Therefore, the corrections for the high-redshift bins are less precise than those for the lower-redshift bins. We do not down-sample the spec-$z$ galaxies due to the limited sample volume. If a larger sample with tens of millions of spec-$z$ galaxies becomes available in the future, the recovery accuracy can be improved.
It is possible that our correction for the photo-$z$ dilution may still contain some residual bias. To address this, we could introduce a nuisance parameter for the photo-$z$ bias of lens and source samples, similar to what many Stage III weak lensing teams have done.
However, since this paper is mainly a demonstration and our primary goal is not to reach the accuracy and precision of cosmological parameters, we will leave this enhancement for future research.

We have assessed the quality of our shear catalog by testing the accuracy of shear recovery using field distortion signals, as presented in \cite{Zhang2022_decals}. This allows us to detect any systematic errors in shear recovery as a function of the source position on the focal plane, which can be caused by a variety of hardware-related effects, such as charge transfer inefficiency \cite{bouchy09,massey14}, CCD defects, residuals of flat field corrections, tree rings \cite{plazas14,jarvis20}, etc. We identified some systematic shear biases related to tree rings using the source catalog of \cite{Zhou2021}, but not with the catalog of \cite{Zou2019}, which contains two to three times fewer sources. We are still investigating the reason for this. In this work, we choose the latter for the lensing measurement. To be compatible with the 2PCF analysis in this work, the photometric redshifts of the sources are still from \cite{Zhou2021}, which covers more than 99\% of the sources of \cite{Zou2019}. We will have a separate paper discussing the systematics in shear measurement in detail.

Table~\ref{table:sample_info} lists a (residual) multiplicative bias for each source sample, $m$, as one of the potential issues. This bias could lead to a few percent error in the ESD measurements, approximately $(1+m)$. We did not take into account a correction for this as the value is consistent with zero within a 2$\sigma$ level.
With these multiplicative biases, the redshift evolution of $\sigma_8$ would be more pronounced than what is shown in Fig.~\ref{fig:cosmos_parameter} given that $\xi_{\rm gg} \propto b^2_g\sigma^2_8$ and $\xi_{\rm gm} \propto b_g\sigma^2_8$. As other Stage-III weak lensing teams have done, we could incorporate a nuisance parameter to account for this bias. However, such a nuisance parameter would be in complete degeneracy with linear galaxy bias and $\sigma_8$, whose precision is largely determined by the prior. 
We will leave such an improvement to future work.

In the modeling part, we make some simplifications when modeling the magnification bias $\Delta\Sigma^{\rm mag}(\rp)$ in Eq.~\ref{eq:esd_mag}. We ignore the redshift uncertainty in both lens and source samples, and instead approximate them with their mean redshift. Additionally, we use a fixed luminosity slope for the lens samples, rather than allowing it to be a free parameter. These simplifications will have an effect on the estimation of cosmological parameters, particularly at high redshifts.

\section{Conclusions}
\label{sec:summary}

We constrain the cosmological parameters $\Omega_m$ and $\sigma_8$ by combining galaxy clustering and galaxy-galaxy weak lensing. Previous studies have used either angular 2PCFs from photometric lens samples or projected 2PCFs from spectroscopic lens samples for galaxy clustering. Both of these methods are accurate. However, interpreting angular 2PCFs can be difficult due to the degeneracy caused by different modes, which can lead to similar angles. On the other hand, projected 2PCF measurements from spectroscopic redshift surveys are limited by Poisson noise and cosmic variance, particularly at high redshifts.

In this study, we use the DESI Legacy Imaging Surveys Data Release 9, which covers an area of approximately 10000 square degrees and spans the redshift range of [0.1, 0.9], to recover the intrinsic projected 2PCFs $w_{\rm p}(r_{\rm p})$ and measure the galaxy-galaxy weak lensing signals (ESDs, $\Delta\Sigma(\rp)$) for lens samples. We assume that the photometric redshift of our lens galaxies follows a Gaussian PDF with respect to their true redshift and divide them into seven approximately volume-limited samples. We then model the observables with the minimal bias model in the context of a flat $\Lambda$CDM cosmology, using conservative scale cuts of $r_{\rm p} > 8$ and $12 ~h^{-1}{\rm Mpc}$ for $w_{\rm p}(r_{\rm p})$ and $\Delta\Sigma(\rp)$, respectively. The extensive lens and source galaxies provide us with highly precise measurements of clustering and weak lensing, allowing us to obtain a tight constraint of the cosmological parameters for each lens sample.

The main results can be summarized as follows.

\begin{itemize}

\item To make sure that the photo-$z$ of lens galaxies follows a Gaussian distribution, we first created a sample of bright and red galaxies whose photo-$z$ have been thoroughly tested and modeled as Gaussian in \cite{Zhou2021}. We then compared the distributions of galaxy photo-$z$ to their true redshifts in weighted spectroscopic subsamples that were similar to the photometric sample in terms of color and magnitude distribution. We found that the distributions were very similar to Gaussian distributions, although there was a small bias of approximately 0.001 (Fig.~\ref{fig:photoz_err}). Finally, the success in predicting projected 2PCFs at different projection lengths (Fig.~\ref{fig:model_prediction}) and the consistency in photo-$z$ uncertainties derived from clustering and from the color-redshift relation (Fig.~\ref{fig:sigmaz_compare}) further confirmed our assumption.

\item The wealth of lenses and sources makes it possible to make highly accurate measurements of clustering and lensing signals (as shown in Figure~\ref{fig:obsVSmodel}). This should help enable a strong constraint on cosmological parameters.

\item After marginalizing out the photo-$z$ uncertainty and the linear galaxy bias, 
both $\Omega_m$ and $\sigma_8$ show a strong evolution with redshift, 
while the derived parameter--structure growth parameter $S_8$--- shows a mild redshift enhancement dependence with low significance (see Fig.~\ref{fig:cosmos_parameter}). 
This trend reveals the strong degeneracy between $\Omega_m$ and $\sigma_8$ and emphasizes the greater robustness of $S_8$ against systematics.  
The constraining power is found to be limited by the accuracy of the lensing measurements, as the precision of galaxy clustering measurements is close to the one percent level (see Fig.~\ref{fig:obsVSmodel}).

\item We compare our $S_8$ constraints with those obtained from Planck 2018 CMB in Fig.~\ref{fig:cosmos_parameter}. At lower redshifts, our constraints agree with Planck, while at higher redshifts, our best-fit values tend to be higher than those of other studies, though with larger errors. When combining (non-overlapping) low redshift samples to fit a constant value of $S_8$, we find that our $S_8 = 0.84\pm0.02$ is in line with the Planck within 1$\sigma$. This value is much higher than the constraints from the latest $3\times2$pt analysis of weak lensing surveys by $\sim 2-5\sigma$.

\item Our method has been shown to be in broad agreement with the canonical values of $S_8$. However, further improvements in measurements and modeling (as discussed in Section~\ref{sec:diss}) are necessary to increase the precision and accuracy of our final constraints.

\end{itemize}

It is currently quite difficult to draw a firm conclusion 
that the redshift-evolution of $\sigma_8$, $\Omega_m$ and $S_8$ were not caused by the systematics. However, if these trends turn out to be true, it might indicate that the underlying cosmology is not LCDM. 
To verify this trend, one approach is to address the systematics as other weak lensing teams have done. An alternative is to use galaxy groups/clusters (e.g. constructed from DR9 \cite{Yang2021}) for the analysis (H.Xu et al. in prep.), which in general suffers less from various systematics. The photo-$z$ of groups with a few member galaxies is usually more reliable than that of individual galaxies. Additionally, the linear bias of galaxy groups/clusters can be obtained directly from the halo mass function (e.g. \cite{WangJiaqi2022}), which will help to strengthen the constraints.

\Acknowledgements{HX thanks Mehdi Rezaie and Edmond Chaussidon for 
their detailed answer to the questions on imaging systematics mitigation.
The authors are grateful to the anonymous referees for their invaluable feedback which significantly enhanced the quality of this paper.
The authors thank Zheng Zheng for careful proofreading.
This work made use of the Gravity Supercomputer at the 
Department of Astronomy, Shanghai Jiao Tong University.
This work is supported by the National Key Basic Research and Development Program of China (No. 2018YFA0404504), the National Science Foundation of China (grant Nos. 11833005, 11890691, 11890692, 11533006, 11621303, 12073017), Shanghai Natural Science Foundation, Grant No. 15ZR1446700 and 111 project No. B20019. We acknowledge the science research grants from the China Manned Space Project with NOs. CMS-CSST-2021-A01, CMS-CSST-2021-A02. 
L.P.F acknowledges the support from National Science Foundation of China grant No. 11933002, the Innovation Program 2019-01-07-00-02-E00032 of Shanghai Municipal Education Commission, and the science research grants from the China Manned Space Project with NO. CMS-CSST- 2021-A01.
The Photometric Redshifts for the Legacy Surveys (PRLS) catalog used in this paper was produced thanks to funding from the U.S. Department of Energy Office of Science,  Office of High Energy Physics via grant DE-SC0007914.
{The DESI Legacy Imaging Surveys consist of three individual and complementary projects: the Dark Energy Camera Legacy Survey (DECaLS), the Beijing-Arizona Sky Survey (BASS), and the Mayall z-band Legacy Survey (MzLS). DECaLS, BASS and MzLS together include data obtained, respectively, at the Blanco telescope, Cerro Tololo Inter-American Observatory, NSF’s NOIRLab; the Bok telescope, Steward Observatory, University of Arizona; and the Mayall telescope, Kitt Peak National Observatory, NOIRLab. NOIRLab is operated by the Association of Universities for Research in Astronomy (AURA) under a cooperative agreement with the National Science Foundation. Pipeline processing and analyses of the data were supported by NOIRLab and the Lawrence Berkeley National Laboratory. Legacy Surveys also uses data products from the Near-Earth Object Wide-field Infrared Survey Explorer (NEOWISE), a project of the Jet Propulsion Laboratory/California Institute of Technology, funded by the National Aeronautics and Space Administration. Legacy Surveys was supported by: the Director, Office of Science, Office of High Energy Physics of the U.S. Department of Energy; the National Energy Research Scientific Computing Center, a DOE Office of Science User Facility; the U.S. National Science Foundation, Division of Astronomical Sciences; the National Astronomical Observatories of China, the Chinese Academy of Sciences and the Chinese National Natural Science Foundation. LBNL is managed by the Regents of the University of California under contract to the U.S. Department of Energy. The complete acknowledgments can be found at https://www.legacysurvey.org/.
Any opinions, findings, and conclusions or recommendations expressed in this material are those of the author(s) and do not necessarily reflect the views of the U. S. National Science Foundation, the U. S. Department of Energy, or any of the listed funding agencies.
The DESI collaboration are honored to be permitted to conduct scientific research on Iolkam Du’ag (Kitt Peak), a mountain with particular significance to the Tohono O’odham Nation.}
}

\InterestConflict{The authors declare that they have no conflict of interest.}


\bibliographystyle{scpma-zycai} 
\bibliography{ms}

\begin{appendix}




\renewcommand{\thesection}{Appendix}

\section{}

\subsection{Red lens samples}
\label{subsec:appendix1}

\begin{figure*}[!t]
\centering
\includegraphics[width=\textwidth]{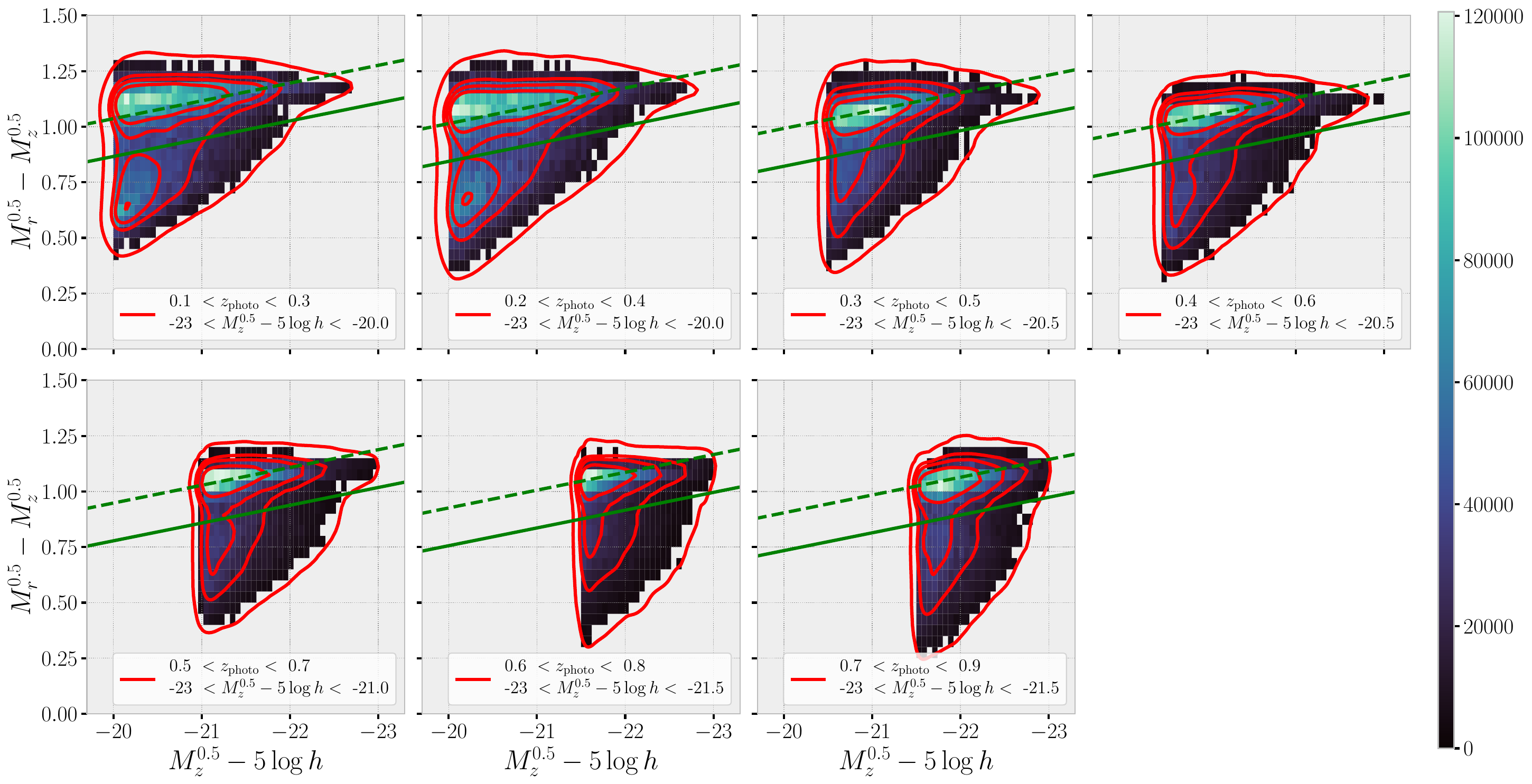}
\caption{Color-magnitude diagram in luminoisty-bin volume-limited samples shown in Fig.~\ref{fig:vol_limit}.
The redshift and luminosity information is indicated in the legend of each panel. 
The $y$-axis is the K-corrected (to $z = 0.5$) color $M^{0.5}_r-M^{0.5}_z$ and the $x$-axis $z$-band absolute magnitude $M^{0.5}_z-5\log h$.
To expedite plotting, 
we randomly draw 50,000 galaxies from each volume-limited sample.
The 2D histograms show 
the galaxy number counts per magnitude per color, i.e.
${\rm d^2N_{gal}}$/d$(M^{0.5}_z-5\log h)/{\rm d}(M^{0.5}_r-M^{0.5}_{\rm z})$. 
The red contours offer a cleaner view of density variation in color-magnitude diagram. 
The green solid lines (Eq.\ref{eq:color}) divide the luminosity
samples into red and blue subsamples, accounting for the both dependence on  luminosity and redshift.
We estimate the slope of green solid lines from the red sequence (green dashed line) of redshift bin 
$0.4 < z_{\rm photo} < 0.6$. 
}
\label{fig:CMD}
\end{figure*}

In section~\ref{subsec:sample_photoz}, we describe the construction of luminosity-bin volume-limited lens samples.
In this appendix, we describe how to select out red galaxies that make our lens samples.

Following the color division practices at low redshift, 
we plot the color-magnitude diagram for luminoisty-bin volume-limited samples, shown in Fig.~\ref{fig:CMD}. 
In all samples, the contours present two distinct galaxy populations: 
a tight red sequence and a lose blue cloud.
This feature allows us to divide the galaxy into red and blue subsamples 
by the following equation,
\begin{equation}
\label{eq:color}
M^{0.5}_r-M^{0.5}_z = -0.8 - 0.08\times(M^{0.5}_z-5\log h)-0.22\times(z-0.5)
\end{equation}
where $M^{0.5}_r-5 \log h$ is the 
K-corrected (to $z=0.5$) $r$-band absolute magnitude. 
The color $M^{0.5}_r-M^{0.5}_z$ seems to work well 
for the division of blue and red subsamples.
It might due to the fact that the 4000 $\AA$ break moves to 
$r$-band at $z\sim0.5$, where the K-correction is performed. 
Note that at relatively high redshift, dust would 
make galaxies appear red, though they might be star-forming. 
We would ignore this complication for this . 


\subsection{Systematic test in Galaxy-Galaxy lensing measurement}\label{subsec:appendix2}
\begin{figure*}[!t]
\centering
\includegraphics[width=\textwidth]{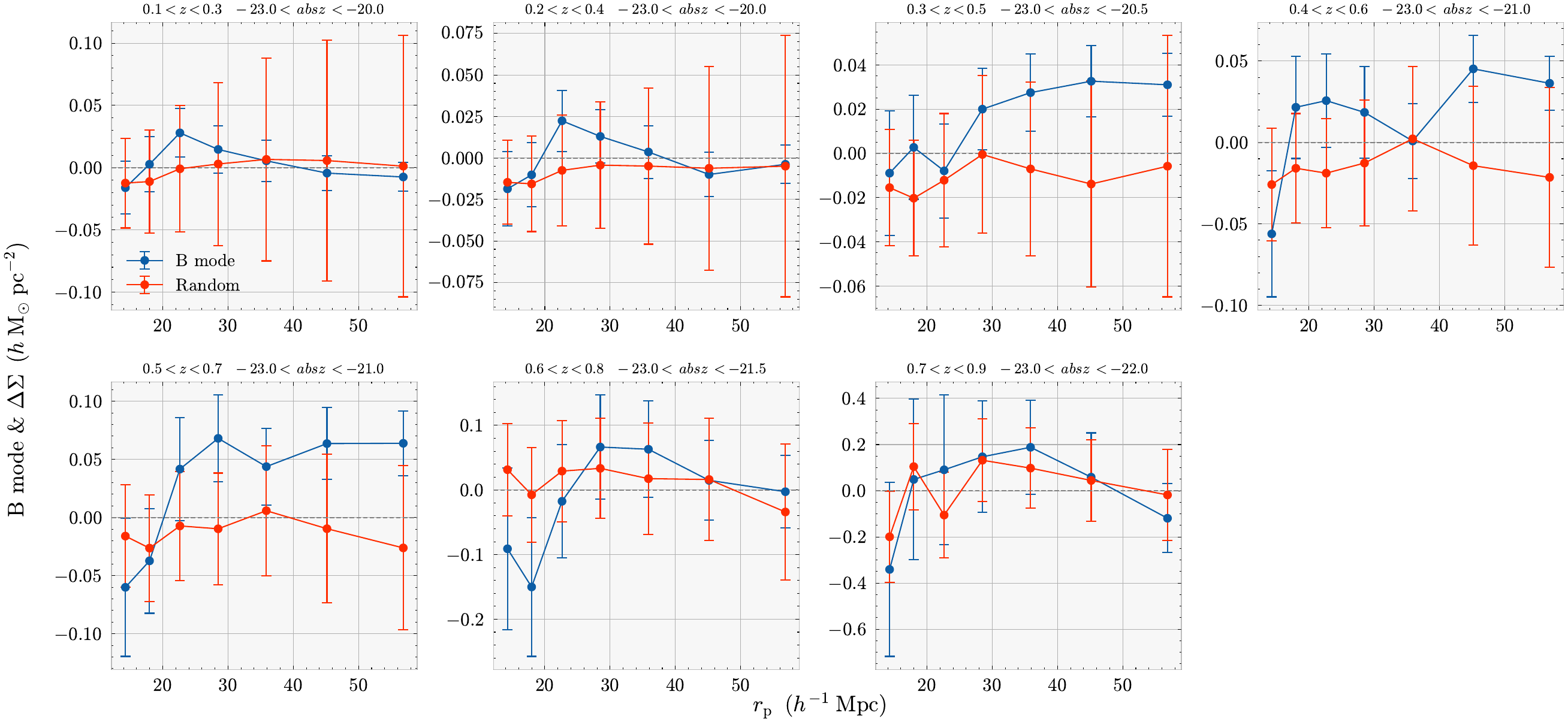}
\caption{The B mode (blue) from lens sample and the ESDs measured around the random (red).
}
\label{fig:lens_sys}
\end{figure*}

In Fig.\ref{fig:lens_sys}, we present the B mode of the lens sample and the ESDs measured from the random catalog as the validation of our ESD measurements. All results are consistent with zero within a 2$\sigma$ level. 
The scatter of B mode of high redshift samples is larger than that of the low redshift samples, due to the photo-$z$ quality.

\begin{figure*}[!t]
\centering
\includegraphics[width=\textwidth]{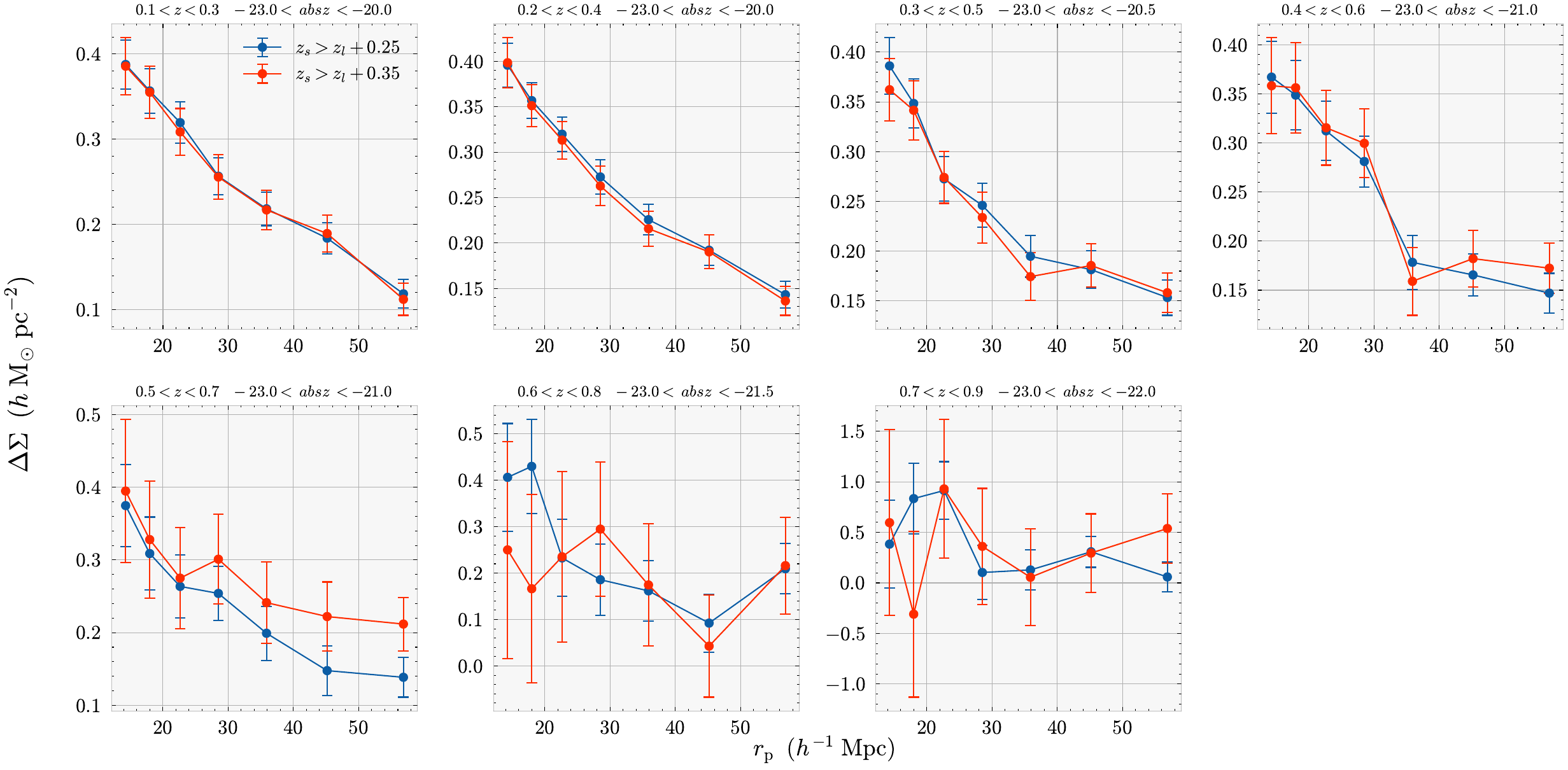}
\caption{The comparsion of ESD measurements with a different lens-sources photo-$z$ separation. The ESDs measured with a larger lens-sources photo-$z$ separation ($\Delta z=0.35$, red) is statistically consistent with the fiducial measurements ($\Delta z = 0.25$, blue) within 1$\sigma$, excpet for the bin $0.5 < z_{\rm photo} < 0.7$.
}
\label{fig:lens_sys_dz_cut}
\end{figure*}

To see how sensitive our ESD measurements to the photo-$z$ quality of source galaxies, 
we choose a larger lens-source photo-$z$ separation, $\Delta z=0.35$, to re-measure the ESDs of lens samples, shown in Fig.~\ref{fig:lens_sys_dz_cut}. We find that ESDs measured with such a larger lens-source photo-$z$ separation do not behave statistically different compared to the fidual measurements (except in the bin $0.5 < z_{\rm photo} < 0.7$).
The reason is probably that the photo-$z$ scatter of source galaxies is much smaller than the fiducial separation ($\Delta z=0.25$), which suggests that our fiducial cut is probably safe to prevent the dilution from the foreground or member galaxies. 


\end{appendix}

\end{multicols}
\end{document}